\pdfoutput=1

\documentclass[fleqn,usenatbib]{mnras}

\usepackage{newtxtext,newtxmath}

\usepackage[T1]{fontenc}

\DeclareRobustCommand{\VAN}[3]{#2}
\let\VANthebibliography\thebibliography
\def\thebibliography{\DeclareRobustCommand{\VAN}[3]{##3}\VANthebibliography}

\usepackage{subcaption} 
\usepackage{gensymb}    
\usepackage{graphicx}	
\usepackage{amsmath}	
\usepackage{xcolor, soul} 

\sethlcolor{yellow}

\title[Feature Guided Training and Rotational Standardisation]{Feature Guided Training and Rotational Standardisation for the Morphological Classification of Radio Galaxies}

\author[K. Brand et al.]{
Kevin Brand$^{1}$\thanks{E-mail: kevinbrand99@gmail.com},
Trienko L. Grobler$^{1}$\thanks{E-mail: tlgrobler@sun.ac.za},
Waldo Kleynhans$^{2}$,
Mattia Vaccari$^{3,4,5}$,
Matthew Prescott$^{4}$,
\newauthor
Burger Becker$^{1}$\\
\\
$^{1}$Computer Science Department, Stellenbosch University, Cnr Banghoek Road \& Joubert Street, Stellenbosch, 7600, South Africa\\
$^{2}$Department of Electrical Electronic and Computer Engineering, University of Pretoria, Pretoria, South Africa\\
$^{3}$Inter-University Institute for Data Intensive Astronomy, Department of Astronomy, University of Cape Town, 7701 Rondebosch, Cape Town, South Africa\\
$^{4}$Inter-University Institute for Data Intensive Astronomy, Department of Physics and Astronomy, University of the Western Cape,\\Robert Sobukwe Road, 7535 Bellville, Cape Town, South Africa\\
$^{5}$INAF - Istituto di Radioastronomia, via Gobetti 101, 40129 Bologna, Italy
}

\date{Accepted XXX. Received YYY; in original form ZZZ}

\pubyear{2023}

\begin{document}
\label{firstpage}
\pagerange{\pageref{firstpage}--\pageref{lastpage}}
\maketitle

\begin{abstract}
State-of-the-art radio observatories produce large amounts of data which can be used to study the properties of radio galaxies. However, with this rapid increase in data volume, it has become unrealistic to manually process all of the incoming data, which in turn led to the development of automated approaches for data processing tasks, such as morphological classification. Deep learning plays a crucial role in this automation process and it has been shown that convolutional neural networks (CNNs) can deliver good performance in the morphological classification of radio galaxies. This paper investigates two adaptations to the application of these CNNs for radio galaxy classification. The first adaptation consists of using principal component analysis (PCA) during preprocessing to align the galaxies' principal components with the axes of the coordinate system, which will normalize the orientation of the galaxies. This adaptation led to a significant improvement in the classification accuracy of the CNNs and decreased the average time required to train the models. The second adaptation consists of guiding the CNN to look for specific features within the samples in an attempt to utilize domain knowledge to improve the training process. It was found that this adaptation generally leads to a stabler training process and in certain instances reduced overfitting within the network, as well as the number of epochs required for training.
\end{abstract}

\begin{keywords}
radio continuum:galaxies -- methods: data analysis -- methods: statistical -- techniques: image processing
\end{keywords}



\section{Introduction}
Most, if not all, massive galaxies have a supermassive black hole at their centre. Active galaxies are galaxies that have an Active Galactic Nucleus (AGN), that is formed by the accretion of particles onto their central supermassive black hole \citep{Padovani2017}. Radio galaxies are active galaxies that emit in the radio part of the electromagnetic spectrum, typically via the synchrotron emission produced from electrons that are accelerated by magnetic fields, present in galaxies \citep{Stacy2003}. This emission can most effectively be observed by making use of radio interferometers \citep[][]{Hardcastle2020}.

The morphological classification of radio galaxies consists of grouping them into classes according to their shapes. This classification process and the definition of a classification system describing the wide variety of observed radio galaxies play a vital role in answering other questions that are fundamental to astronomy, such as how galaxies and galaxy clusters form and evolve over cosmic time \citep[][]{Hardcastle2020}.

Two important morphological classes, namely class I and II Fanaroff-Riley galaxies (FRI and FRII galaxies), were first identified by \citet{Fanaroff1974}. To differentiate between these classes, the Fanaroff-Riley ratio (FR ratio) was defined as the ratio of the distance between their `hot spots' (radio bright areas observed on either side of a galaxy's core) to the total distance between the edges of the radio source. Galaxies with an FR ratio less than $0.5$ have bright regions close to the galaxy's core and are classified as FRI galaxies, whilst galaxies with an FR ratio greater than $0.5$ are edge-brightened and are classified as FRII galaxies.

Some radio galaxies have a more complex warped shape, i.e. the two radio jets emanating from their centres are not aligned to form a straight line but are rather at an angle with each other. These radio galaxies are known as bent-tail radio galaxies and they can be separated into two further classes depending on the angle between their jets, namely narrow-angle tailed \citep[NAT,][]{Rudnick1976} and wide-angle tailed galaxies \citep[WAT,][]{Owen1976}. For the rest of this paper, these subclasses will be grouped together and referred to as bent galaxies.

The final class of radio galaxies that we will consider in this paper is the FR0 class, also known as the compact class. This class was proposed for the classification of a subset of galaxies that do not fit into the original FR dichotomy very well. These galaxies have a very bright region surrounding the core, but as explained by \citet{Baldi2015}, they have a much more compact morphology. Galaxies with other, more complex morphologies fall outside the scope of this paper.

With the development of new facilities, such as the Australian SKA Pathfinder \citep[ASKAP,][]{Johnston2008}, the MeerKAT radio telescope \citep[][]{Jonas2009}, the Low-Frequency Array \citep[LOFAR,][]{Haarlem2013}, and the Murchison Widefield Array \citep[MWA,][]{Beardsley2019}, it has become possible to conduct surveys that will detect much larger numbers of radio galaxies than was previously possible. As the rate of radio source detection grows, it becomes increasingly difficult to manually analyze and classify each source. One approach to solving this problem would be to make use of crowd-sourcing, where large groups of individuals voluntarily classify each source \citep[][]{Banfield2015}. However, this solution relies on the availability of enough individuals that can generate reliable labels, which might not be feasible as the volume of data increases, especially once the construction of the Square Kilometre Array \citep[SKA,][]{Dewdney2009, Dewdney2013} has been completed.

An alternative approach has been gaining traction in recent years. This approach consists of automating the analysis and labelling of sources and is much more robust to an influx in the volume of data that is available. Machine learning is a useful tool in this automation process and has been shown to be quite effective \citep[e.g.][]{DeLaCalleja2004, Alhassan2018, Hocking2018}.

Initially most of the machine learning techniques that were applied were shallow learning techniques, such as decision trees \citep[][]{Proctor2003, Proctor2006} and support vector machines \citep[][]{Sadeghi2021}, which achieved impressive results with respect to the classification of radio galaxies. However, as the field of deep learning continues its rapid growth, a clear shift has been observed towards the use of deep neural networks.

To better understand why this shift is sensible, it is important to understand what some of the limitations are of the shallow techniques and why these limitations are relevant to radio galaxy data. One such limitation is that shallow learning algorithms tend to underperform when compared to deep learning techniques on various high-dimensional datasets \citep[][]{LeCun2015}. Seeing as radio galaxy datasets tend to consist of relatively large images, this can be problematic. Furthermore, when applying shallow machine learning to high-dimensional datasets, a common approach is to reduce the dimensionality by extracting informative features. However, this approach has certain drawbacks. By only training the shallow models on the extracted features, we are introducing bias into the models and preventing them from finding informative features that we might not be aware of \citep[][]{Janiesch2021}. Deep learning has also been applied with great success in applications that utilize large corpora of data \citep[][]{Chen2014}, whilst shallow learning algorithms have been shown to be inefficient in comparison \citep[][]{Najafabadi2015}. Thus, it is expected that deep learning will play a crucial role in extracting information from future large-scale surveys.

With respect to the morphological classification of radio galaxies, \citet{Aniyan2017} found that convolutional neural networks can be very effective. They constructed three CNNs that each played the role of a binary classifier that simply had to differentiate between two classes. The predictions from these binary classifiers were then combined to make a final prediction. This model showed that CNNs could achieve similar performance to manual morphological classification, but that they could do it quicker.

Since then, various new CNNs have been created to perform morphological classification on radio galaxies. Some examples of these CNNs include the single multi-class CNN created by \citet{Alhassan2018} to classify radio galaxies as compact, bent, FRI or FRII galaxies, a multi-class CNN trained on the first dataset from the Radio Galaxy Zoo project \citep[][]{Banfield2015} to differentiate between compact and extended sources \citep[][]{Lukic2018}, CNNs used for the cross-identification of host galaxies \citep[][]{Alger2018}, an augmented Fast Region-based CNN \citep[Fast R-CNN][]{Girshick2015} that was applied to combinations of corresponding radio and infrared images by \citep[][]{Wu2018} to simultaneously find and classify radio sources. \citet{Ma2019} made use of a large number of unlabelled samples to pretrain a convolutional autoencoder (CAE), which is a type of neural network that learns how to encode images into lower dimensional representations. The weights of the encoder section of this network were then fine-tuned, using a set of labelled samples to perform the morphological classification of radio galaxies.

Recently, \citet{Bowles2021} made use of an attention-gating mechanism\footnote{Attention-gating mechanisms work similarly to human attention mechanisms by identifying and focusing on the features that are the most informative for the given task.} instead of fully-connected layers in a CNN to show that one could use considerably fewer parameters to achieve performance that is on par with the CNNs previously applied to morphological classification. Another important development was the development of group-equivariant CNNs for radio galaxy classification by \citet{Scaife2021}, who explained that conventional CNNs are not equivariant to certain isometries, such as rotations and reflections, which contribute to intra-class variability. Thus, these CNNs might not be able to accurately classify a sample during inference if its orientation differs from the samples observed during training, because the rotation will have an effect on the outputs of the convolutional layers and will thus affect the classification. \citet{Scaife2021} addressed this problem by making use of a subspace of conventional convolution kernels that is equivariant to rotations. The use of these kernels led to an improvement in the performance of the CNN, which serves as additional proof that rotations in samples can have an effect on the performance of CNNs.

\subsection{Addressing rotational variations in radio sources}
Making use of group-equivariant CNNs is not the only way to address rotational variations in radio sources. One can also augment the training data by applying rotations to the training samples, thus increasing the number of training samples and making it possible for CNNs to learn what classes look like at various orientations. This approach is widely adopted in the literature \citep[][]{Becker2021, Bowles2021, Alhassan2018, Lukic2018, Aniyan2017, Ma2019}, however it has certain flaws. \citet{Scaife2021} noted that by using data augmentations to address rotational variations, one runs the risk of learning identical kernels for different orientations, which is a waste of some of the computational power of the CNN. Furthermore, they also explain that invariance gained through data augmentation does not guarantee invariance during inference. Aside from the points mentioned by \citet{Scaife2021}, one should also consider that data augmentation can lead to a dramatic increase in the number of samples used during training, which will tend to lead to much longer training times.

Bearing this in mind, this paper investigates a third approach that addresses the rotational variations in radio sources, which attempts to standardise the rotation of all of the radio sources as a preprocessing step. The first step in this process is to construct a matrix that contains the 2D coordinates of the pixels that belong to the galaxy. Principal component analysis (PCA) is then applied to this matrix to determine the principal components of the galaxy. By aligning these principal components with the main axes, we effectively standardise the rotation of the radio sources. A similar approach was applied to optical galaxy images by \citet{DeLaCalleja2004}, but we specifically investigate the effectiveness of this preprocessing step on radio sources and compare it to the commonly used approach of rotational data augmentation.

\subsection{Incorporation of domain knowledge during learning}
The patterns/features that are deemed to be important by neural networks during training are generally unconstrained, which makes it possible for them to arrive at non-optimal solutions \citep[][]{Bader2008, Lawrence1998} or to make decisions that violate known physical constraints within the problem domain \citep[][]{Chen2021, Daw2021}. These networks can also overfit on noise or inconsequential patterns within the training data, but due to the ``black box'' nature of neural networks \citep[][]{Olden2002, Seidel2019} it is very difficult to verify whether the network has learned sensible patterns and to determine what those patterns might be \citep[][]{Seidel2019, Woods2019}. This also means that neural networks cannot truly contribute to the development of our understanding of the problem domain \citep[][]{Olden2002}.

These problems have led to the development of a variety of approaches that attempt to address the ``black box'' nature of neural networks. Some of these approaches train neural networks and then develop an understanding of their decision-making process by extracting symbolic rules from the trained models \citep[][]{Huynh2011, Seidel2019}. The use of symbolic rules lies at the core of knowledge-based neural networks \citep[KBNN,][]{Fu1995, Kolman2008}.

To better understand KBNNs, one should first consider that there are two paradigms of learning, theoretical learning and learning from practical experience. Expert systems follow the theoretical paradigm of learning and utilize expert knowledge of the problem domain during the decision-making process, whilst empirical learning follows the practical paradigm and learns how to solve a task by training on a large number of representative samples from the problem domain \citep[][]{Towell1994}.

Both expert systems and empirical learning systems have been able to achieve incredible results on real-world problems \citep[][]{Goethe1995, Rahman1996, Sun2008}. However, each system has its disadvantages. Expert systems require a large investment of both time and resources to construct complete domain theories. This leads to the use of incomplete theories, which could be detrimental in certain scenarios \citep[][]{Mitchell1986}. Conversely, systems that make use of empirical learning have no domain knowledge, which leads to different problems. For example, without domain knowledge, the system is not aware of any constraints that determine whether results are viable and thus there are no guarantees that they will produce viable solutions \citep[][]{Mitchell1986}. The lack of domain knowledge also means that the performance of the system is entirely dependent on whether the training data is informative with respect to the problem domain. Thus, poor performance is all but guaranteed if the quality of the training data is poor \citep[][]{Schank1986}.

A combination of these two systems might be able to address some of these disadvantages, which is exactly what KBNNs attempt to do. KBNNs combine these two systems by incorporating domain theories into a neural network in the form of symbolic rules. They then make use of empirical learning algorithms to fine-tune the weights of the neural network. This approach has some distinct benefits. For example, the use of empirical learning algorithms and domain samples will refine any incomplete or incorrect domain theories which will lead to better performance. Conversely, if empirical learning leads to results that violate theoretical constraints, domain theories can be used to constrain the network to only produce viable results. Considering these benefits, it should come as no surprise that KBNNs have been shown to be incredibly effective and that they tend to generalize better than both expert and empirical learning systems \citep[][]{Towell1994}.

The implementation details of KBNNs are not relevant to this paper and will not be discussed. However, it is important to note that these networks contain some neurons that are not dependent on domain theory. \citet{Towell1994} showed that the addition of these neurons provides neural networks with the necessary flexibility to refine incomplete or inaccurate domain theories and to address some of the side effects that they cause. The addition of these neurons is relevant to this paper, because this technique will also be used in some of our guided architectures.

Theory-guided data science (TGDS) is a paradigm of data science that one can argue is a continuation of the work done for KBNNs. The goal of TGDS is to produce models that do not exclusively make use of either data or theoretical knowledge, but instead utilize both \citep[][]{Karpatne2017}. This goal is achieved by making use of a wide variety of techniques that introduce domain knowledge into models that are traditionally purely data-driven. One such technique consists of using Lagrangian multipliers to enforce theoretical constraints in the loss functions of models that make use of error-based learning. Other techniques include the use of theoretical knowledge to efficiently initialize parameters, fine-tune model outputs and select probabilistic models that are similar to the known distribution of the problem domain \citep[][]{Karpatne2017}.

A technique that is commonly used in TGDS, which will also be used in this paper, is feature engineering. This technique consists of manipulating and combining existing features to create new features that are expected to be informative when classifying samples. Such features are selected by making use of prior knowledge of the domain and the task at hand \citep[][]{Nargesian2017}. The features that are engineered can be very simple, such as the ratio between two other features \citep[][]{Heaton2016}, or they can be much more complex, such as the results of statistical transformations that were applied to the data. If good features are selected, it has been shown that feature engineering can lead to considerable improvements in the performance of models \citep[][]{Mulaudzi2021, Yu2010}.

Based on the results achieved by both KBNNs and TGDS, this paper investigates whether it would be useful to guide neural networks to look for specific features during training. This approach was first tested on a toy example, where a shallow neural network was constructed to emulate the behaviour of the XOR binary operator. Once it had been determined that such a guided neural network was viable, guided CNNs were constructed and applied to the morphological classification of radio galaxies. The performance of the guided CNNs was evaluated and compared to the performance of unguided CNNs with a similar architecture to determine whether there were any considerable benefits that justify guiding neural networks during training.

Our guided neural networks should not be mistaken with Feature-guided Denoising Convolutional Neural Networks \citep[FDCNN][]{Dong2021}. FDCNNs make use of an explainable artificial intelligence technique, guided backpropagation, to identify and extract important features in ultrasound images. These features are used to improve the performance of a denoising neural network. There are no significant similarities between FDCNNs and our guided networks, other than their name.

Section~\ref{sec:data} of this paper presents the datasets that were used, the preprocessing steps that were applied to the data, as well as the process that was followed to extract features for the training of the guided neural networks. This is followed by an explanation of the neural network architectures in Section~\ref{sec:arch}. The analytic process that was used to evaluate and compare the rotational preprocessing steps and the guided networks is discussed in Section~\ref{sec:analysis}. We then present and discuss the results of our experiments in Section~\ref{sec:results}. Finally, we draw our conclusions in Section~\ref{sec:conclusion}.

\section{Datasets and Preprocessing Procedures}
\label{sec:data}
This section discusses the datasets that were used in this paper, the steps involved in the preprocessing procedures, as well as the feature extraction process. Section~\ref{sec:dataset} presents the datasets and Section~\ref{sec:preprocessing} discusses the various preprocessing techniques that were applied to the radio galaxy images. Finally, Section \ref{sec:feature_extract} explains how features were extracted from these images.

\subsection{Datasets}
\label{sec:dataset}
\subsubsection{XOR binary operator}
The XOR binary operator is quite a simple operator. It is also known as the exclusive or operator, which means that it only returns true if one of two binary inputs is equal to one. This operator was used in this paper as a simple example to test whether a certain approach to guiding neural networks was feasible or not. The XOR dataset is very small and only consists of four samples that cover the entire domain. These samples can be seen in \autoref{tab:xor_data}.

\begin{table}
	\centering
	\caption{XOR dataset}
    \label{tab:xor_data}
	\begin{tabular}{ccc}
		\hline
		Input 1 & Input 2 & Output \\
		\hline
		0 & 0 & 0\\
		0 & 1 & 1\\
		1 & 0 & 1\\
		1 & 1 & 0\\
		\hline
	\end{tabular}
\end{table}

\subsubsection{Radio galaxy dataset}
Our radio galaxy dataset was assembled as follows. We first gathered positional and morphological information from the following well-documented catalogues extensively used in the literature: CoNFIG \citep{2008MNRAS.390..819G,2010MNRAS.404.1719G}, GROUPS \citep{2011ApJS..194...31P}, FR0CAT \citep{2018A&A...609A...1B}, FRICAT \citep{2017A&A...598A..49C}, FRIICAT \citep{2017A&A...601A..81C} and WATCAT \citep{2019A&A...626A...8M}. We then downloaded 300 by 300 pixels cutouts from the FIRST survey \citep{1995ApJ...450..559B,2015ApJ...801...26H} in FITS format using the SkyView online tool \citep{1998IAUS..179..465M}. We then confirmed that the morphology of each radio galaxy as indicated in the different input catalogues was reliable when inspected at the depth and the resolution of the FIRST survey via visual inspection of all radio galaxies by at least three team members. Objects for which consensus was not achieved were collaboratively re-checked. Where consensus could not be achieved after the re-check, the radio galaxy was dropped from our sample. Our resulting sample, which we called FIRST Radio Galaxy Morphology Reference Catalogue (FRGMRC), consists of a total 960 sources. The breakdown in different source classes is shown in Table~\ref{tab:cat}. The FRGMRC and the supporting FIRST fits cutouts used for our work is publicly available at \url{https://doi.org/10.5281/zenodo.7645530}.

\begin{table}
	\centering
	\caption{Number of radio sources from different classes within our sample.}
    \label{tab:cat}
	\begin{tabular}{cc}
		\hline
		Class & Number \\
		\hline
  		Compact & 208 \\
		FRI & 182 \\
	    FRII & 357 \\
		Bent & 213 \\
		Total & 960 \\
		\hline
	\end{tabular}
\end{table}

\subsection{Radio Galaxy Preprocessing}
In this section we will discuss the various preprocessing techniques that were used in this paper. We will provide an overview of what these techniques consist of and what purpose they serve. The details of when each preprocessing step was used are delayed until the discussion of the investigative procedure in Section \ref{sec:invest}.

\label{sec:preprocessing}
\subsubsection{Normalization}
Normalization consists of scaling all pixel values within an image to fall within the range $[0, 1]$. This preprocessing step is always applied to the radio galaxy images in this paper, irrespective of what they will be used for, because the pixel ranges tend to differ between images, which leads to an unnecessary increase in the intra-class variances. Normalization is achieved by applying \autoref{eq:normalize} to each pixel in an image. In this equation, $X$ represents the entire image and $x_{ij}$ represents the individual pixels at row $i$ and column $j$.

\begin{equation}
    x_{ij} = \frac{x_{ij} - \min{(X)}}{\max{(X)} - \min{(X)}}
	\label{eq:normalize}
\end{equation}

\subsubsection{Thresholding}
Thresholding is a technique that is frequently used to extract important pixels from an image by removing noise and background pixels. This technique is especially important in this paper, because it improves the quality of the features that are extracted from the images and it assists with determining the principal components of the galaxies.

We use a slightly different thresholding approach in this paper than is commonly used in the literature due to our rotational standardisation algorithm's sensitivity to thresholding artefacts. The first step in our approach is to identify how much noise is present in the image. This is done by constructing a histogram of pixel values for each image. The number of bins that are populated with a large number of pixels provides an indication of the amount of noise that is present in the image. If there is noise present in the image there will be a wider range of pixel values, which will lead to more histogram bins being populated.

If there is very little noise present in the image we can simply use a static value to threshold the image. Alternatively, if there is noise present, we make use of a large quantile as a threshold to extract the brightest galaxy pixels. We then make use of morphological dilations to `grow' these bright pixels to form regions that contain all of the galaxy pixels. We also make use of morphological operations to find small artefacts in the background and to remove them.

In cases where the noise is severe, we create a new matrix, where each element represents the number of pixel values in the immediate neighbourhoods of the corresponding pixel in the image. We can exploit this information to extract the galaxy pixels, because the pixels along the borders of the galaxy will have background pixels, noise pixels and galaxy pixels in their neighbourhood, which means that the pixel values in their neighbourhoods will be much more diverse. Thus, if we extract the elements with a large range of neighbouring pixel values, we will be able to extract most of the pixels near the border of the radio galaxy. We then apply morphological hole filling, which will fill holes that are surrounded by the pixels that we have extracted. In doing so, we will be able to also extract the galaxy pixels that are not close to the border of the galaxy.

Our approach sometimes includes a small number of pixels from the regions surrounding the galaxy, which might lead to slight distortions of the shape of the galaxy. However, it is very successful in removing thresholding artefacts, which is more important for our algorithms. For more details about our thresholding algorithm, the reader is referred to the Github repository\footnote{The repository is available at \url{https://doi.org/10.5281/zenodo.7785574}}.

\subsubsection{Rotational data augmentation}
\label{sec:rotation_aug}
As was mentioned earlier, an approach that is widely used to address rotational variations in radio galaxy data is to augment the training data by applying rotations to each of the training samples. This approach is also implemented in this paper to allow for a comparison with the method of rotational standardisation.

Each sample in the FRGMRC dataset was rotated in 60 degree intervals, from 0 to 300 degrees, which creates six copies of each sample, but with different orientations. This makes it possible for the CNNs to learn what each class will look like at different orientations.

Ideally, one would use smaller rotational intervals to ensure that the CNN could recognize the classes at any possible orientation, but this is likely to lead to a significant waste of time and computational resources, because the CNN might not learn much from samples that are too similar with regards to their orientation.  

One of the metrics which will be used to evaluate the performance of rotational data augmentation is the amount of time that it takes to train a CNN whilst using this approach. It was determined that with small rotational intervals, data augmentation leads to near-perfect accuracy on the samples in the testing data, which means a large portion of the training time might have been wasted. Thus, a rotational interval was selected that led to a slight reduction in the accuracy of the CNN, because this would ensure that time was not wasted on unnecessary training.

\subsubsection{Rotational standardisation}
\label{sec:rotation_std}
To understand the approach that we use to standardise the rotation of the radio galaxy images, it is vital to first understand principal component analysis (PCA). The origins of PCA are unclear, but it can be traced back as far as 1829. However, the form of PCA that is commonly used today is attributed to the work done by \citet{Hotelling1933}.

PCA is commonly used as a technique for dimensionality reduction. As explained by \citet{Rao1964}, PCA consists of finding orthogonal vectors that describe the data the best. To do so, it is necessary to find the vectors along which the variance in the data is maximized. These vectors were established to be the eigenvectors of the covariance matrix of the data and are also known as the principal components of the dataset. Thus, the first principal component will be the direction in which the variance is the largest, the second principal component will be the direction with the second largest variance etc. To reduce the dimensionality of the data to $q$ dimensions, each sample can be projected onto the first $q$ principal components, which represent the $q$ vectors along which the variance in the data is the largest and will thus lead to the smallest loss of information.

In this paper, we will not use PCA to reduce the dimensionality of our data, but will instead use it to identify the vectors that represent the directions of maximum spread in our radio galaxies. We do so by first normalizing and thresholding each sample to identify the pixels that belong to the radio galaxies. The quality of PCA is correlated to the quality of this thresholding, so any improvements in the thresholding algorithm will lead to an improvement in the results of PCA.

Once we have identified the galaxy pixels, we construct a $2 \times P$ matrix, $X$, where $P$ is the number of pixels that were extracted. Each column in $X$ represents the coordinates of one of the pixels. The next step consists of centring the galaxy at the origin of the coordinate system. This is done by subtracting the mean of each row in the matrix from the elements in that row. We then make use of the popular NumPy library \citep[][]{Harris2020} to calculate the singular value decomposition (SVD) of the centred matrix. This decomposition is shown in \autoref{eq:svd}. For more information regarding the SVD, the reader is referred to \citet{Klema1980}

\begin{equation}
    X = U\Sigma V^{T}
	\label{eq:svd}
\end{equation}

In this decomposition $U$ is a $2 \times 2$ orthogonal matrix, $\Sigma$ is a $2 \times P$ diagonal matrix and $V^{T}$ is a $P \times P$ orthogonal matrix. We will only need the $U$ matrix, because its columns contain the eigenvectors of the covariance matrix of $X$, which we know are also the principal components of $X$. We also know that these principal components maximize the variance of our data.

Seeing as our data is the coordinates of the galaxy pixels, these principal components will maximize the spread of the galaxy pixels. Thus, if we rotate our coordinate system's axes to align with these principal components, we should be able to remove any rotation from our image. 

One way to achieve this goal would be to premultiply the transpose of $U$ with the coordinate of each pixel in our image. This would provide us with the corresponding coordinates in a coordinate system that is aligned with the principal components. However, this approach would be unnecessarily complicated and very inefficient.

Instead, we start by determining whether $U$ is the identity matrix. If it is, we can determine that no rotation is necessary to align our coordinate system with the principal components. If $U$ is not the identity matrix, we need to calculate the angle with which we have to rotate our image. We know that premultiplication by the transpose of the $U$ matrix will lead to a rotation in our data and potentially a reflection. Reflections are possible, because the sign of an eigenvector can be changed and the result will still be an eigenvector. Thus, the sign of any of the columns of $U$ can be changed and the column will still represent a principal component. This means that $U^T$ has to be a combination of an arbitrary number of reflection matrices and a rotation matrix.

Consider that a rotation matrix has the form shown in either \autoref{eq:rotate} or \autoref{eq:rotate2}, depending on the direction of the rotation. These matrices are not in their conventional form, because the flipped $y$-axis in computer graphics changes the direction of rotation.

\begin{equation}
    R^{\circlearrowright} = \begin{bmatrix}
        \cos{(\theta)} & -\sin{(\theta)}\\
        \sin{(\theta)} & \cos{(\theta)}
    \end{bmatrix}
	\label{eq:rotate}
\end{equation}

\begin{equation}
    R^{\circlearrowleft} = \begin{bmatrix}
        \cos{(\theta)} & \sin{(\theta)}\\
        -\sin{(\theta)} & \cos{(\theta)}
    \end{bmatrix}
	\label{eq:rotate2}
\end{equation}

By utilizing this knowledge, we should be able to extract the angle, $\theta$, from $U$. We do so by calculating the $\arcsin$ of the first element of the second column of $U^T$. We know that this element might have been multiplied by $-1$, but seeing as $-\sin{(\theta)} = \sin{(-\theta)}$, this should not affect the absolute value of $\theta$.

Once we have calculated the rotation angle, we need to determine whether we should perform a clockwise or an anti-clockwise rotation. Before we can determine the direction of the rotation, we will first need to determine whether the sign of any of the columns has been altered.

We know that the sign of $\theta$ does not affect the result of the cosine function. Thus, we can use this result to establish whether the sign of one of the columns have been changed. We do so by calculating $\cos{(\theta)}$ and comparing the result to the first element of the first column and the second element of the second column of $U$. If the sign of the elements in $U$ differs from the sign of our calculated value, we can assume that the sign of the corresponding column has been flipped. We reverse this reflection by multiplying that column with $-1$.

This process should address any reflections, meaning that we should be left with only a rotation matrix. To determine whether we need to perform a clockwise or anti-clockwise rotation, we have to determine which of the sine elements in $U^T$ is the negative element. We do so by dividing each sine element with $\sin{(\theta)}$ and inspecting the sign of the results \footnote{In the scenario where the sine elements have the same sign and the cosine elements are zero we will not be able to determine which column's sign has been changed. However, in this scenario the rotation angle has to be 90\degree, which means the direction of the rotation is not important.}. If the first element of the second column has a negative sign, we know that the rotation matrix is in the form shown in \autoref{eq:rotate}, otherwise it is in the form shown in \autoref{eq:rotate2}.

\begin{figure*}
    \centering
	\begin{subfigure}{.42\linewidth}
    \includegraphics[width=\linewidth]{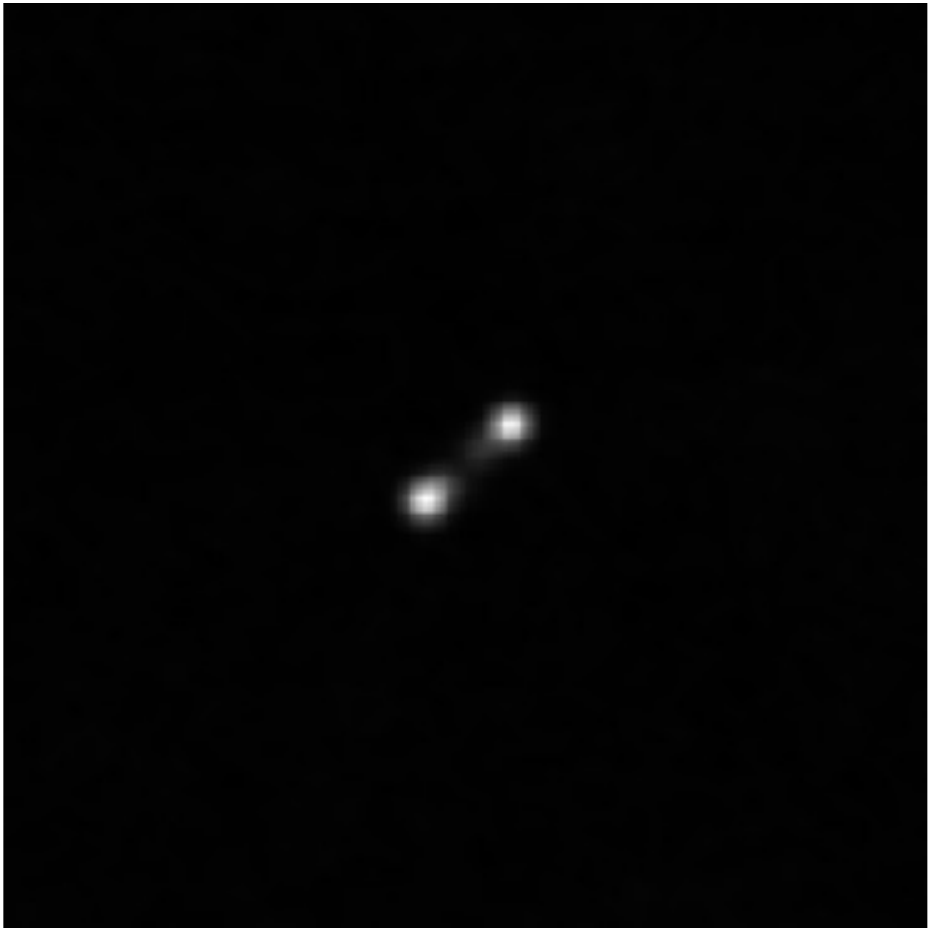}
    \caption{Original image}\label{fig:pre_rotate}
    \end{subfigure}
    \begin{subfigure}{.42\linewidth}
    \includegraphics[width=\linewidth]{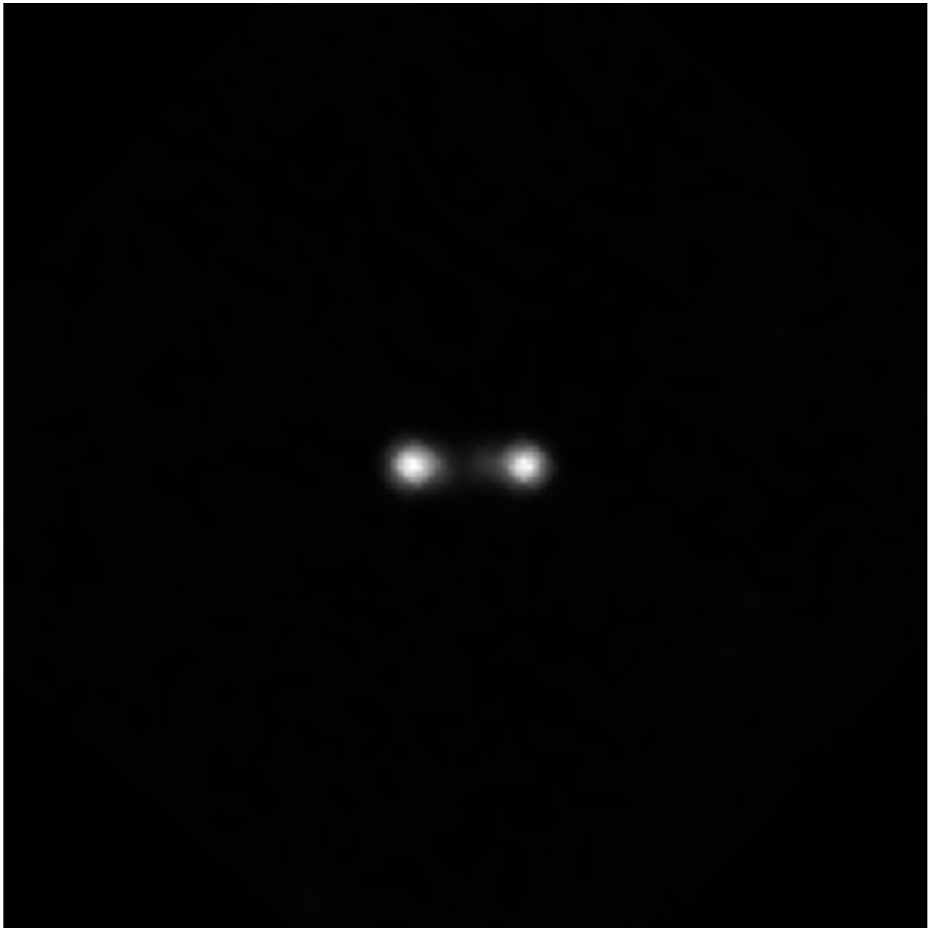}
    \caption{Image after standardising the orientation}\label{fig:post_rotate}
    \end{subfigure}
    \caption{Results of applying the rotational standardisation algorithm, which leads to a 41.68\degree\ clockwise rotation.}.
    \label{fig:rotate}
\end{figure*}

We tested our method of standardising the orientation of the radio galaxies, by determining whether we could use the rotation angle that we calculated and the assumptions that we made to reconstruct the $U^T$ matrix. We could do so successfully for all of the samples in our training data. For an example of the results that this algorithm can achieve, the reader is referred to \autoref{fig:rotate}.

\subsection{Radio Galaxy Feature Extraction}
\label{sec:feature_extract}
We want to guide the CNNs to look for specific features in the samples that we consider to be informative. This section will discuss the features that we selected, as well as how we extracted them. We extracted two sets of these features. One set was exclusively extracted by algorithms, whilst in the other, two of the features were extracted manually. This was done to determine whether either approach of extracting features resulted in significant improvements in the performance of the guided networks.

We selected four features that we deemed to be informative with respect to the morphological classification of the radio galaxies. These features include the FR ratio, whether the galaxy is bent or not, the number of so-called `bright spots' that are present in the galaxy, which we will refer to as cores for the rest of the paper, and the ratio of the size of the cores to the total size of the galaxy.

Before extracting any features, the images had to be normalized and thresholded. Once this had been done, the first step consisted of calculating the standard deviation of the galaxy pixels. The cores of the galaxy were then extracted by using a threshold that depended on this standard deviation. A number of thresholds were used during experimentation to find the thresholds that delivered the best results. These thresholds can be seen in \autoref{tab:cores_thresh}.

\begin{table}
	\centering
	\caption{Thresholds used for core extraction.}
    \label{tab:cores_thresh}
	\begin{tabular}{cl}
		\hline
		Standard deviation range & Threshold \\
		\hline
		$0.2 < \sigma$ & 80th quantile\\
		$0.13 < \sigma <= 0.2$ & 93rd quantile\\
		$\sigma <= 0.13$ & 98th quantile\\
		\hline
	\end{tabular}
\end{table}

The number of pixels larger than the threshold are then counted. This count is divided by the total number of pixels in the galaxy, which leads to the first feature that we want to extract, the ratio of pixels in the cores to pixels in the galaxy. This feature indicates whether the cores dominate the galaxy, which is especially prevalent in compact galaxies.

Once this feature had been extracted we could count the number of cores in the galaxy by counting the number of connected components in the thresholded image. Whilst counting the connected components, we also recorded the coordinates of the pixel with the maximum value in each component. We used these coordinates to represent the center of each component. If there are two cores we can simply use the distance between them to represent the inter-core distance in the FR ratio. If there are more than two cores we calculated the distance between each core and its nearest neighbour and used the average distance to represent the inter-core distance. If there was only one core, it could mean that we are dealing with a compact source, but there is also the possibility that it is an FRI source where the distance between the cores is so small that they look like one core. Thus, we calculate the distance from one side of this core to the other side and use a third of this distance to represent the inter-core distance. This is only an estimate of what the distance would be if cores are overlapping, but it is accurate enough to indicate that the cores are very close to one another.

Before we can continue with feature extraction, it is necessary to determine whether there was a lot of noise present in the sample. The reader might remember that if there is a lot of noise in a sample, the thresholding algorithm tends to include some background pixels or noise that is connected to the galaxy, which complicates the process of identifying potential curves in the galaxy's shape.

Thus, if we detect a lot of noise during thresholding, we use the standard deviation and mean of the extracted galaxy pixels to calculate the z-score. The z-score represents how many standard deviations each pixel is from the mean. We want to get rid of any background pixels that were mistakenly extracted from the original image. These background pixels are generally much darker than the rest of the pixels, which means that they will have a considerably larger z-score than the rest of the pixels. Thus, we only keep pixels with a z-score that is smaller than the median z-score. This might seem to be a very aggressive threshold, but it was found to deliver good results for the next steps of the feature extraction process.

Once we had removed any potential background pixels from the image, we used an approach that is very similar to the approach from Section \ref{sec:rotation_std}. The only difference was that we multiplied the $U$ matrix from the SVD directly with the matrix that contains the coordinates of the galaxy pixels. This multiplication produced the coordinates that the pixels would have if we rotated them to align the principal component with the axes. By calculating the difference between the minimum and maximum column value in our new coordinate system, we can determine the approximate distance from one side of the galaxy to the other. The inter-core distance can then be divided by this distance to calculate the FR ratio.

The final step is to determine whether the galaxy is bent. We do this by using the distance between the minimum and maximum column that we just calculated, as well as the distance between the minimum and maximum row. If the galaxy is not bent, one of these distances should be considerably larger than the other, but if the galaxy is bent these distances should be closer to one another.

Unfortunately, some factors complicate the process of identifying curves in practice. These factors include the imperfect identification of galaxy pixels, the small angle between tails in NATs, the similar vertical and horizontal dimensions in certain compact, FRI and FRII galaxies, etc.

To select good thresholds for these values, we took into consideration that bent galaxies are much more likely to have a curve in the galaxy's body, whilst curves in the bodies of other classes of radio galaxies should be much rarer. Thus, we selected thresholds that would separate most of the bent galaxies from the other galaxies. There were no thresholds that could perfectly achieve this separation, but we do not view this as a problem, seeing as we want the feature to indicate whether a curve is present in the galaxy. We do not want it to become a class label for bent-tailed galaxies.

The hyperparameters used in this process are specific to this dataset. They can be heavily optimized, because this algorithm will not be applied to unseen samples, meaning that there is no risk of overfitting. However, if this approach is applied to a new dataset, it might be necessary to tune these hyperparameters for the new samples.

We also manually determined the number of cores in each galaxy and whether the galaxy was curved in any way. These features were not extracted by an astronomer and thus simply indicate the number of regions in an image that were considered to be cores and whether it seemed like a galaxy was curved in any way. The manually extracted features were used in the additional set of features that were mentioned earlier to determine whether manually extracting features leads to any significant improvements in performance.

\section{Architectures}
\label{sec:arch}
For this paper, a few guided, as well as unguided neural networks were constructed. These networks consist of a standard convolutional neural network that simply classifies samples, an auxiliary convolutional neural network that is used to extract the selected features from samples and three guided convolutional neural networks. These guided networks are referred to as the wide, multiheaded and merged neural networks and they are all guided to use information from the extracted features to a certain extent when classifying samples.

This section will explain the architectures of the CNNs that were applied to the FRGMRC data as well as the process of hyperparameter selection. The full details of these architectures can be found in the \href{https://doi.org/10.5281/zenodo.7785574}{Github repository}. The architectures of the networks that were applied to the XOR data can be seen in Appendix~\ref{appendix:XOR_arch}.

The reader will notice that the architectures in this section contain dense blocks and convolutional blocks. Each convolutional block contains two convolutional layers, followed by a max pooling layer. The dense blocks each consist of a dense layer, which is followed by a dropout layer.

These architectures have not been extensively optimized. This was done purposefully, because we want to be able to identify whether the changes that we make to the network architectures affect overfitting and model performance. If the models were already fine-tuned to the point where there is no overfitting present in the networks and the network performs perfectly, it would be difficult to determine whether our changes affect the models' performance.

Thus, we are of the opinion that it is more important to ensure that the hyperparameters and the general structure remain as similar as possible in the various networks to ensure that there are no factors that could affect network performance, other than the factors that we are investigating.

\subsection{Standard convolutional neural network}
\begin{figure*}
    \centering
    \includegraphics[width=.7\linewidth]{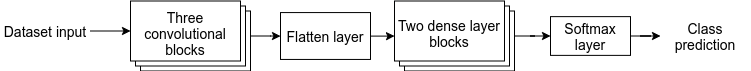}
    \caption{Structure of the SCNNs.}
    \label{fig:tradnet}
\end{figure*}

As the name suggests, this network simply represents a standard unguided convolutional neural network as can be seen in \autoref{fig:tradnet}. It was used as a baseline with which to compare the performance of the guided convolutional neural networks. The standard convolutional neural network (SCNN) was also used to compare our rotational standardisation technique to rotational augmentation.

\subsection{Auxiliary convolutional neural network}
\begin{figure*}
    \centering
    \includegraphics[width=.7\linewidth]{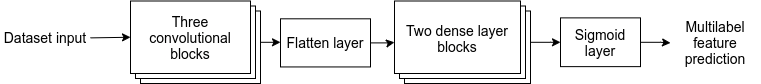}
    \caption{Structure of the ACNNs.}
    \label{fig:auxnet}
\end{figure*}

A representation of the auxiliary convolutional neural network (ACNN) architecture can be seen in \autoref{fig:auxnet}. From this figure it can be seen that the ACNN architecture is very similar to that of the SCNN. The only difference is that the output layer in the ACNN makes use of a sigmoid activation function to produce multilabel feature predictions, whilst the SCNN makes use of a softmax activation function to make a single class prediction. This network is used to engineer new features for the datasets, by determining the presence of the predetermined features within each of the given samples.

\subsection{Wide convolutional neural network}
\begin{figure*}
    \centering
    \includegraphics[width=.7\linewidth]{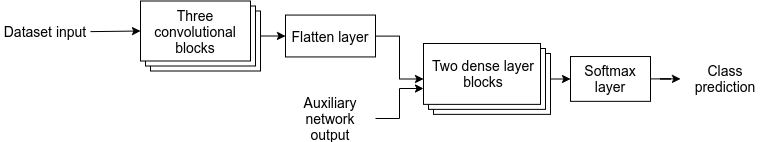}
    \caption{Structure of the WCNNs.}
    \label{fig:widenet}
\end{figure*}

The wide convolutional neural network (WCNN) was based on the neural network architecture presented by \citet{Cheng2016}. Similar architectures have also been applied to radio galaxy data \citep[e.g.][]{Tang2022}. \autoref{fig:widenet} shows a representation of the modified architecture used for the WCNN in this paper.

These networks have the benefit that they can learn from multiple inputs in parallel. In this paper, the network specifically receives additional inputs just before the first dense layer blocks. These additional inputs are the features engineered by the ACNNs. They are fed into the WCNNs after the convolutional layers, seeing as the convolutional layers are meant for image data and will probably not be very useful if applied to the engineered features.

These networks are the first guided neural networks and were used to establish whether feature engineering using domain knowledge could improve the training process in any way.

\subsection{Multiheaded convolutional neural network}
The multiheaded convolutional neural networks (MhCNN) are based on an existing neural network architecture that has been used for a wide variety of applications. These networks generally have a shared base, which branches out into a number of heads that each performs a specific task. For an example of these multiheaded neural networks, the reader is referred to the work done by \citet{Li2019}.

\begin{figure*}
    \centering
    \includegraphics[width=.7\linewidth]{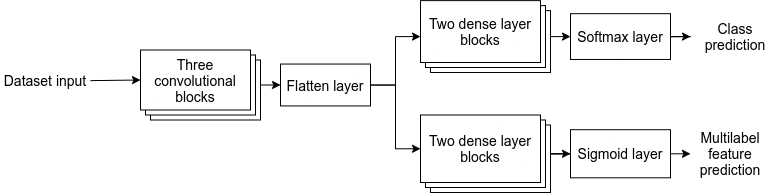}
    \caption{Structure of the MhCNNs.}
    \label{fig:multinet}
\end{figure*}

The structure of the multiheaded networks used in this paper is shown in \autoref{fig:multinet}. As can be seen, the MhCNNs split up into two separate heads after the convolutional layers. One head is similar to the ACNN, whilst the other head is similar to the SCNN.

At first glance, it might not be apparent that MhCNNs are guided neural networks. However, the reader should note that the two heads share multiple convolutional blocks. The optimization algorithm will thus adjust the weights according to an aggregate loss function which uses the loss functions of both heads. This means that the auxiliary head will have an effect on parameters in the convolutional layers, which will in turn have an effect on the main head. Thus, during the training process, the auxiliary head will still be able to guide the main head, albeit indirectly. Ideally, the shared convolutional layers will find patterns that are beneficial to both the auxiliary and main heads, which could speed up training.

After training, MhCNNs provide the extracted features as an additional output, which can then be used for other tasks. One such task could be to make use of these features to validate the class assigned to each sample. For example if the network did not detect any curves in the body of a sample, but classified it as a bent-tailed galaxy, one might need to double check the result. Such a network could also be used to learn additional labels for the samples, such as the number of peaks and components in each sample, similar to what was done for CLARAN \citep[][]{Wu2018}.

\subsection{Merged convolutional neural network}
The merged convolutional neural networks (MCNNs) are a continuation of the fundamental concepts used in the multiheaded and wide convolutional neural networks. These MCNNs also make use of engineered features to guide the training process, but they construct these features within the hidden layers of the network. They then feed these engineered features into the output layer.

A variant of the merged neural network was first constructed for the XOR binary operator, seeing as the network structure could be kept simple\footnote{Refer to Appendix~\ref{appendix:XOR_arch}}. This made it possible to determine the updated terms from first principles. Doing so proved that backpropogation and stochastic gradient descent were still applicable to networks that made use of the outputs of hidden neurons in their loss function.

\begin{figure*}
    \centering
    \includegraphics[width=.7\linewidth]{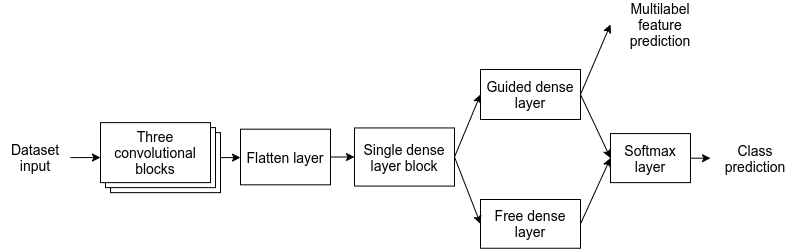}
    \caption{Structure of the MCNNs.}
    \label{fig:mergenet}
\end{figure*}

Once it had been confirmed that it was possible to construct a merged neural network, MCNNs were constructed for the FRGMRC dataset. A representation of the architecture of these networks can be seen in \autoref{fig:mergenet}. As can be seen in this figure, the final dense layer is split into two sections. One section consists of `guided' neurons that are used to extract the selected features from the sample, whilst the other consists of standard `free' neurons that are free to find any patterns in the given sample that will assist in classification.

MCNNs have similar benefits to those of the MhCNNs. They also learn to extract features from the samples, which can be used for other downstream tasks, such as verifying whether the classification labels are sensible or not. They also have the added benefit that the extracted features are fed back into the dense layer that classifies the samples, which means that the extracted features can play a more direct role in the classification process.

\subsection{Hyperparameter selection}
Hyperparameters in machine learning models can drastically change model behaviour and thus influence the quality of the output. Thus, it is important to select good values for them.

Unfortunately, due to the large number of interchangeable, complex elements present in neural networks, they tend to have quite a large number of hyperparameters. This makes it infeasible to explore the entire hyperparameter space in search of good values. Instead, this paper evaluated the networks' performance on a validation set for a few values of each hyperparameter and then selected the values that seem to lead to the best performance. 

A Nadam optimizer with a learning rate of 0.0001 was used for all of the networks in this paper. A grid search was conducted that determined the best learning rate for the SCNNs. The same learning rate was used for the other neural networks, seeing as their convergence speed would be compared.

For the hidden layers, ReLU and ELU activation functions were used. ReLU activation functions were used in the convolutional layers and ELU activation functions were used in the dense layers. With respect to weight initialization, He normal weight initialization was used for all of the dense layers and Xavier uniform initialization was used for the convolutional layers. The bias terms were all initialized to zero.

\section{Empirical analysis}
\label{sec:analysis}
In this section the analytic process will be discussed that was used to compare the various neural networks. Section~ \ref{sec:metrics} will discuss the loss functions and performance metrics that were used. This will be followed by a discussion of the procedure that was used to generate reliable performance evaluations for the rotational preprocessing steps as well as each of the neural network architectures.

\subsection{Loss functions and metrics}
\label{sec:metrics}
To be able to compare the various neural networks, it was necessary to select a few performance metrics. This subsection will discuss these metrics as well as the loss functions that were used.

\subsubsection{Categorical cross-entropy loss}
Categorical cross-entropy takes the probability of a sample belonging to a class and compares it with the true label of that sample. It then computes a logarithmic penalty which corresponds to how far away the prediction probability is from the desired output. This calculation can be seen in \autoref{eq:crossent}. In this equation $N$ refers to the number of samples in the dataset, $M$ refers to the number of classes and $p_{ij}$ is the output from the neural network corresponding to class $j$ and sample $i$. Finally, $t_{ij} = 1$ if sample $i$ belongs to class $j$. Categorical cross-entropy loss is used for models that make a single class prediction. Thus, this loss function was used for the neural networks that classified samples from the FRGMRC dataset to evaluate the class predictions that they made.

\begin{equation}\label{eq:crossent}
    C = -\frac{1}{N}\sum^{N}_{i=1}\sum^{M}_{j=1} t_{ij}\log_2{(p_{ij})}
\end{equation}

\subsubsection{Binary cross-entropy loss}
The TensorFlow implementation of binary cross-entropy loss is very similar to categorical cross-entropy loss, except that it is used for binary classification. The equation used to calculate this loss is very similar to the equation used for categorical cross-entropy loss, as can be seen in \autoref{eq:bincrossent}. The value $N$ still refers to the number of samples, $t_{i}$ represents the true label for sample $i$ and $p_{i}$ represents the predicted label. This loss function was used to evaluate how well neural networks were able to extract the bent feature from the FRGMRC data.

\begin{equation}\label{eq:bincrossent}
    C = -\frac{1}{N}\sum^{N}_{i=1} t_{i}\log_2{(p_{i})} + (1 - t_{i})\log_2{(1 - p_{i})}
\end{equation}

\subsubsection{Mean squared error}
\label{sec:mse}
The mean squared error (MSE) loss function is commonly used to evaluate the performance of neural networks when they are utilized to predict continuous variables. In this paper it was used to evaluate how well neural networks could extract the continuous-valued features from the radio galaxies, such as the FR-ratio and the number of cores.  This loss can be seen in \autoref{eq:MSE}. $N$ represents the number of samples, $t_i$ is the true value of sample $i$ and $p_i$ is the value predicted by the network. 

\begin{equation}\label{eq:MSE}
    C = \frac{1}{N}\sum^{N}_{i=1}(t_i - p_i)^2
\end{equation}

A variant of this loss function was also used to optimize the networks that were trained on the XOR dataset. The only difference was that the entire equation was divided by 2. This has no impact on the optimization process, because the weights that minimize $C$ will also minimize $\frac{C}{2}$. This division is generally excluded in the literature, but it has been used here to simplify the derivative of the loss function.

\subsubsection{Mean absolute error}
The mean absolute error (MAE) is very similar to the mean squared error. The only difference is that it uses the absolute distance between predictions and true values instead of the squared distance. We also make use of this metric to evaluate how well neural networks could extract the continuous-valued features from the radio galaxies. This metric is included, because it makes it easier to interpret how well the neural network can extract the continuous-valued features. The MAE can be seen in \autoref{eq:MAE}. $N$ represents the number of samples, $t_i$ is the true value of sample $i$ and $p_i$ is the value predicted by the network. 

\begin{equation}\label{eq:MAE}
    C = \frac{1}{N}\sum^{N}_{i=1}|t_i - p_i|
\end{equation}

\subsubsection{Combined loss function}
The MhCNNs and the MCNNs have two outputs, the features that were extracted and the class label of the sample. To ensure that all of the outputs are optimized, the loss function has to incorporate at least one component for each output that can provide an empirical evaluation of the quality of the output.

However, not all of these outputs are necessarily equally important. It might be more important to extract certain features, or it might be more important to classify the sample. Thus, weights were used to indicate how important each component was. The weighted loss function can be seen in \autoref{eq:combined_loss}. In this equation, $m$ represents the main component of the loss, the categorical cross-entropy loss, $a$ represents the auxiliary component which indicates how well the networks were able to extract the specified features from the samples and $w_1$ and $w_2$ represent the weights that correspond to each component.

\begin{equation}\label{eq:combined_loss}
    C = w_{1}m + w_{2}a
\end{equation}

\subsubsection{Classification accuracy}
The classification accuracy of a model is very easy to calculate. It is simply the number of correct classifications divided by the number of samples in the given dataset. This metric gives a good indication of the models' classification performance.

\subsubsection{F1 Score}
\label{sec:f1_score}
The F1 score is a slightly more intricate performance metric. To understand the F1 score, it is necessary to first understand what is meant by the precision and recall of a network. Precision is calculated as shown in \autoref{eq:precision}. In this equation $TP$ is the number of true positives and indicates how frequently the model accurately identifies the presence of the positive class. In a binary classification problem, the positive class is the primary class that one is interested in. The opposite is true for the negative class. $FP$ is the number of false positives and indicates how frequently the model falsely identifies the presence of the positive class.

\begin{equation}\label{eq:precision}
    \text{Precision} = \frac{TP}{TP + FP}
\end{equation}

\autoref{eq:recall} shows how the recall of the model is calculated. $FN$ is the number of false negatives and indicates how frequently a model falsely determines that the positive class is not present.

\begin{equation}\label{eq:recall}
    \text{Recall} = \frac{TP}{TP + FN}
\end{equation}

These two metrics are then combined as shown in \autoref{eq:f1} to create the F1 score. For this paper, the F1 score was used as a metric to evaluate the classification performance of the networks on the FRGMRC data. This was necessary, because there is a slight imbalance in the class representations and the F1 score is less sensitive to these imbalances than the classification accuracy is. In this case, the positive class is simply whether a sample belongs to one of the morphological classes.

\begin{equation}\label{eq:f1}
    \text{F1} = 2\times\frac{\text{Precision}*\text{Recall}}{\text{Precision} + \text{Recall}}
\end{equation}

Seeing as the F1 score was applied to problems that were not binary classifications, the calculations had to be adapted. The precision and recall was calculated for each class. The average precision and recall for all of the classes was then used to calculate the overall F1 score. This is commonly known as the macro-average F1 score.

\subsubsection{Overfitting metric}
\label{sec:overfit_metric}
We want to be able to determine whether guided neural networks generalize better than unguided neural networks. To be able to compare these networks we need a metric that provides an indication of how much overfitting is present in the network. One indicator that is commonly used, is the difference between the training loss and the validation loss. If overfitting occurs, the training loss will keep improving, whilst the validation loss might plateau or even become worse.

Before we calculate the difference between the training and validation losses, it is important to acknowledge that the losses that are being optimized in the various networks differ from one another and have different scales. Thus, it would not be fair to compare them directly, because losses with a larger range of values will automatically make it seem like the network is overfitting more severely.

To address this problem, we first record all of the training and validation losses for every training run of a network and then use these losses to formulate an approximation of the range of values that we can expect to observe for that loss. We then use this information to normalize the losses, by using an equation similar to \autoref{eq:normalize}. The only difference is that $X$ will now represent all of the loss values that were recorded and $x$ will refer to one of these loss values.

Once this has been done, all of the losses will have values between zero and one, which makes it possible to draw a fair comparison between the various networks. At this point we can simply calculate the difference between the training loss and the validation loss when training finishes. The larger this difference is, the more overfitting has occurred.

\subsection{Investigation procedure}\label{proc}
Due to the stochastic nature of neural networks, one cannot expect the training process to deliver consistent results. To get a reliable indication of the performance of the various networks, it is necessary to train each network multiple times and to then use statistical measures to evaluate their performance over all of the runs.

Thus, each of the neural networks in this paper were trained and evaluated over twenty independent runs. The performance metrics were recorded for each run. For the radio galaxy networks this was done by making use of the Tensorboard callback in Tensorflow \citep[][]{Abadi2015}. After all of the runs had finished, the logs were used to determine the average performance of the networks.

To establish how well the networks generalize, the FRGMRC dataset was split up into a training, validation and test dataset before the first training run. The sizes of these datasets are given in \autoref{tab:data_split}. The total number of samples in this table does not align with the total number of samples in \autoref{tab:cat}. The difference is due to some compact sources that were added to the FRGMRC dataset after experimentation, as well as a few duplicate sources that were dropped at an advanced stage of the project. Neither of these changes affect our results, because the models were all trained and evaluated on the same set of samples.

\begin{table}
	\centering
	\caption{Number of samples in training, validation and test dataset.}
    \label{tab:data_split}
	\begin{tabular}{cccc}
		\hline
		Training & Validation & Test & Total\\
		\hline
		702 & 87 & 88 & 877\\
		\hline
	\end{tabular}
\end{table}

Recall that we also trained shallow networks for the XOR dataset as a feasibility study. These networks were trained for a maximum of 10000 epochs on the the full dataset. This is acceptable, because the XOR dataset is exhaustive. The radio galaxy CNNs were trained on the training dataset for a maximum of 100 epochs. After each epoch, they were also evaluated on the validation dataset, which indicated how well the models were generalizing to unseen data.

To save some time an early stopping mechanism was used to stop the training of the XOR neural networks when the loss function reached a value that was smaller than 0.01. Early stopping was also used to stop the training of the radio galaxy networks if the validation loss had not improved for 5 epochs. This assisted in addressing the overfitting in the network to a certain extent. Once the training of the radio galaxy networks had concluded, the network's state was reset to the state when the minimum validation loss was achieved. The network was then evaluated on the test dataset.

\subsubsection{Rotational Standardisation Investigation Details}\label{sec:invest}
In this paper we compare the effectiveness of rotational augmentation and rotational standardisation to address rotational variations that are commonly found in the radio galaxy data. In these experiments SCNNs were trained on the FRGMRC dataset after different preprocessing steps were applied.

To evaluate rotational augmentation, each of the training and validation datasets were augmented, using the approach detailed in Section~\ref{sec:rotation_aug}. The test dataset was not augmented, because this would lead to an unfair comparison of the various approaches, since they would effectively be evaluated on different datasets. All of the samples were then normalised, after which a SCNN was trained and evaluated on the processed datasets over twenty training runs.

Rotational standardisation was evaluated in a similar fashion. The only difference was that instead of augmenting the data, the rotation of each of the samples in the training, validation and test datasets was standardised, using the approach that was explained in Section~\ref{sec:rotation_std}.

Finally, we also trained and evaluated an SCNN on a dataset where the only preprocessing that was done consisted of normalising the samples. This SCNN was used as a baseline to determine how important it is to address the rotational variations in a given dataset.

\subsubsection{Feature Guided Training Investigation Details}
To investigate whether there are any benefits in guiding the networks to look for specific features, three guided CNN architectures were evaluated, as well as one unguided architecture, which served as a baseline with which to compare their performance. The guided architectures included WCNNs, MCNNs and MhCNNs.

As mentioned earlier, the MCNNs and MhCNNs make use of weights to balance the contribution of the main and auxiliary outputs to the total loss of the network. To get an indication of the effect that these weights have, three MhCNNs and three MCNNs were trained. Each network made use of different weight combinations, which can be seen in \autoref{tab:weights}. All of the components of the loss functions had a similar order of magnitude, which is why the weights also have a similar order of magnitude.

\begin{table}
	\centering
	\caption{Weights used to balance loss components in radio galaxy networks.}
    \label{tab:weights}
	\begin{tabular}{lll}
		\hline
		Network & Main weight & Auxiliary weight\\
		\hline
		Even MhCNN & 0.5 & 0.5\\
		Main MhCNN & 0.75 & 0.25\\
		Auxiliary MhCNN & 0.25 & 0.75\\
		Even MCNN & 0.5 & 0.5\\
        Main MCNN & 0.75 & 0.25\\
        Auxiliary MCNN & 0.25 & 0.75\\
		\hline
	\end{tabular}
\end{table}

Before training and evaluating these networks, each sample was normalized. No other preprocessing steps were applied. We decided not to threshold the images before training, because during experimentation a SCNN performed considerably worse when trained on thresholded images. The SCNN achieved an average accuracy of $86.2\%$ without thresholding and an average accuracy of $82.05\%$ with thresholding over twenty training runs.

\section{Results}
\label{sec:results}
This section will present and discuss the results from our experiments. The results of our rotational standardisation algorithm will be presented in Section~\ref{sec:rotate_results}. This will be followed by a discussion of the performance of the guided CNNs on the XOR and FRGMRC dataset in Section~\ref{sec:guided_xor_results} and Section~\ref{sec:guided_results} respectively.

\subsection{Rotational Standardisation Results}
\label{sec:rotate_results}
Recall that we trained three SCNNs to evaluate our rotational standardisation algorithm. One was trained on a dataset where no rotational processing was performed, one was trained on a dataset that was augmented with rotated versions of the samples and one was trained on a dataset where our rotational standardisation algorithm was applied as a preprocessing step. The average macro F1, loss and training time of each of these networks can be seen in \autoref{tab:rotate_results}.

\begin{table*}
	\centering
	\caption{Average performance of SCNNs with various rotational preprocessing strategies.}
    \label{tab:rotate_results}
	\begin{tabular}{llll}
		\hline
		Preprocessing & Average Macro F1 & Average Loss & Average Training Time (s)\\
		\hline
		None & $0.876$ & $0.34$ & $116$\\
		Rotation Standardisation & $0.95$ & $0.18$ & $98$\\
		Augmentation & $0.971$ & $0.12$ & $476$\\
		\hline
	\end{tabular}
\end{table*}

This table clearly indicates that approaches that address rotational variations lead to far superior classification results in comparison with approaches that ignore these variations. Such an increase in classification performance validates the claims that rotational variations negatively affect the performance of classifiers.

\begin{figure*}
    \centering
	\begin{subfigure}{.42\linewidth}
    \includegraphics[width=\linewidth]{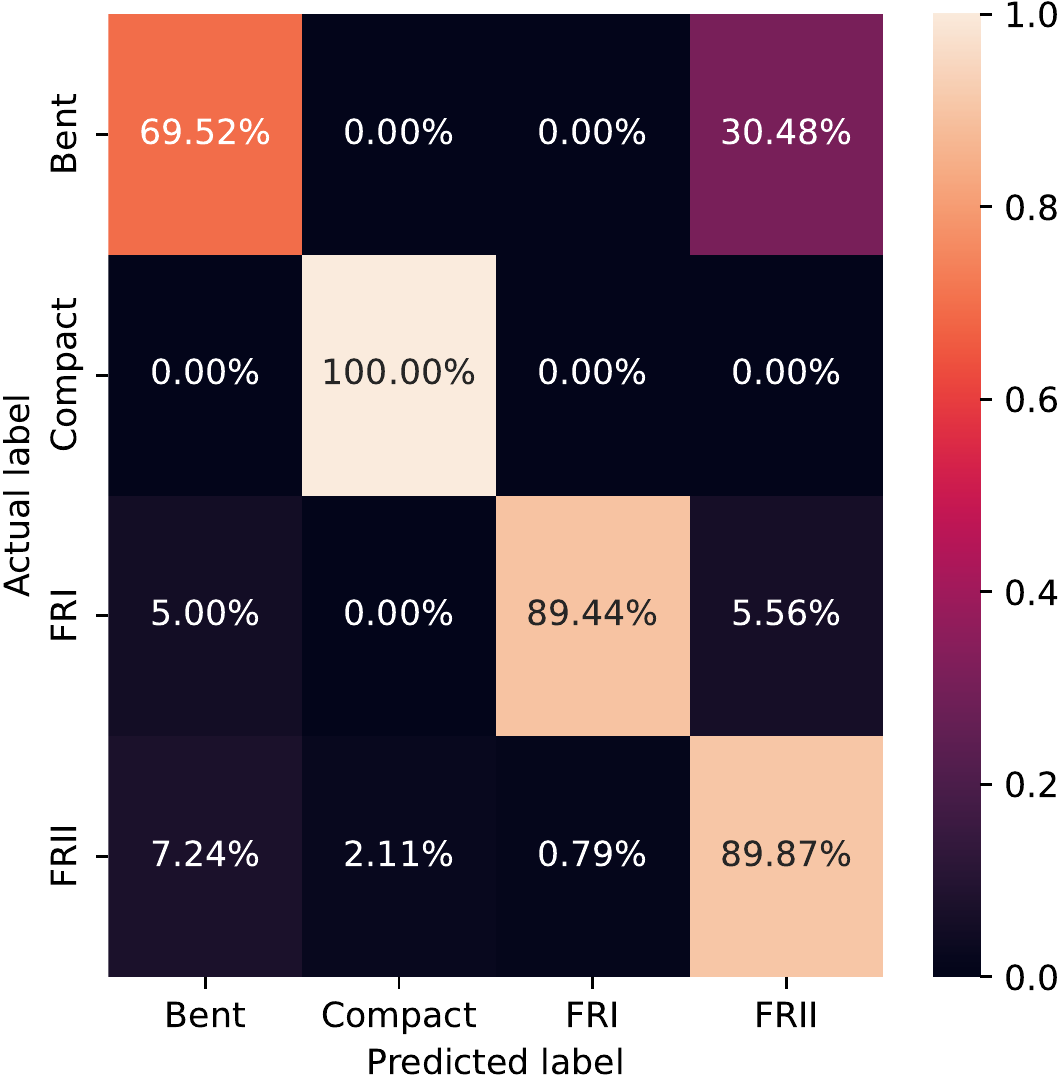}
    \caption{No preprocessing}\label{fig:cm_std}
    \end{subfigure}\hspace{5mm}
    \begin{subfigure}{0.42\linewidth}
    \includegraphics[width=\linewidth]{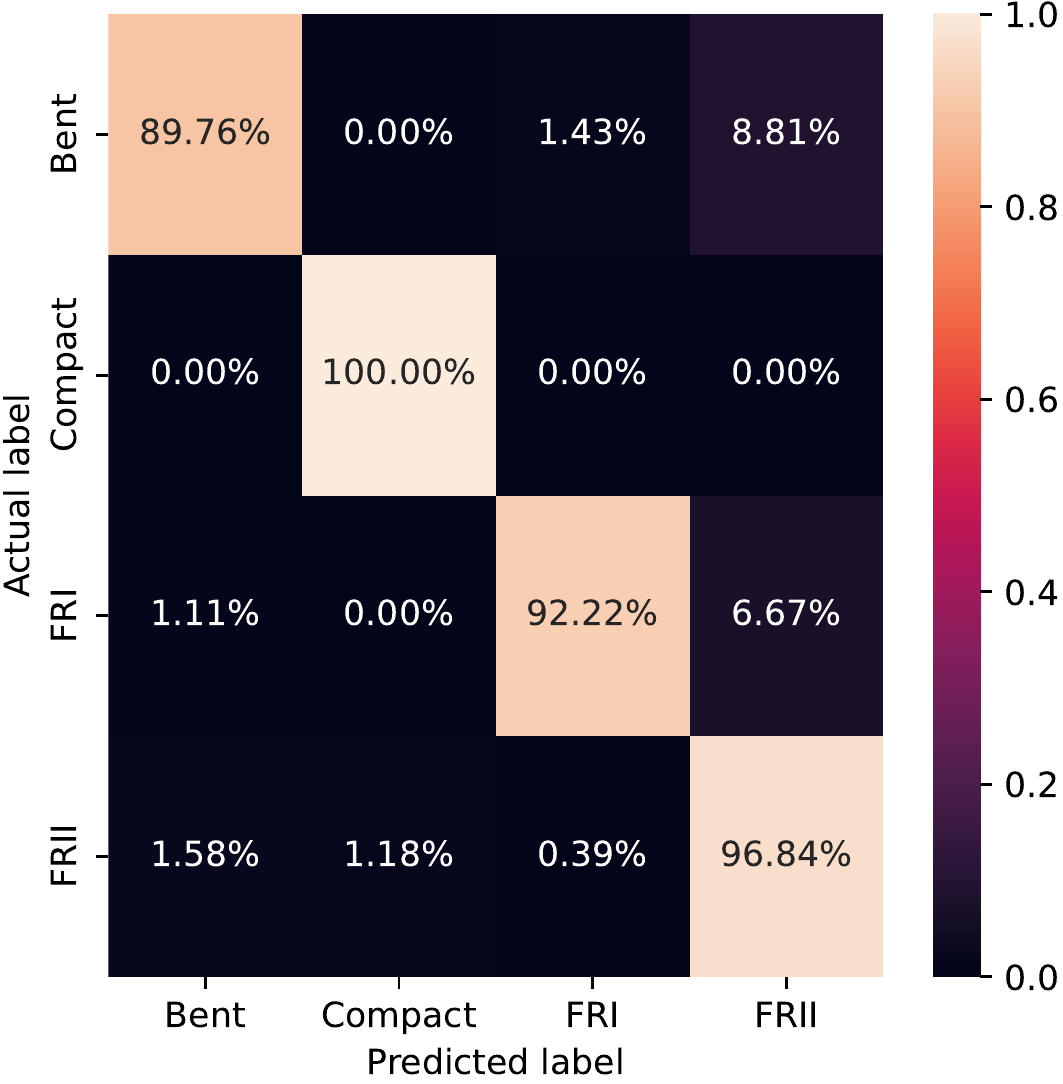}
    \caption{Rotational standardisation}\label{fig:cm_rot}
    \end{subfigure}
    \begin{subfigure}{.42\linewidth}
    \includegraphics[width=\linewidth]{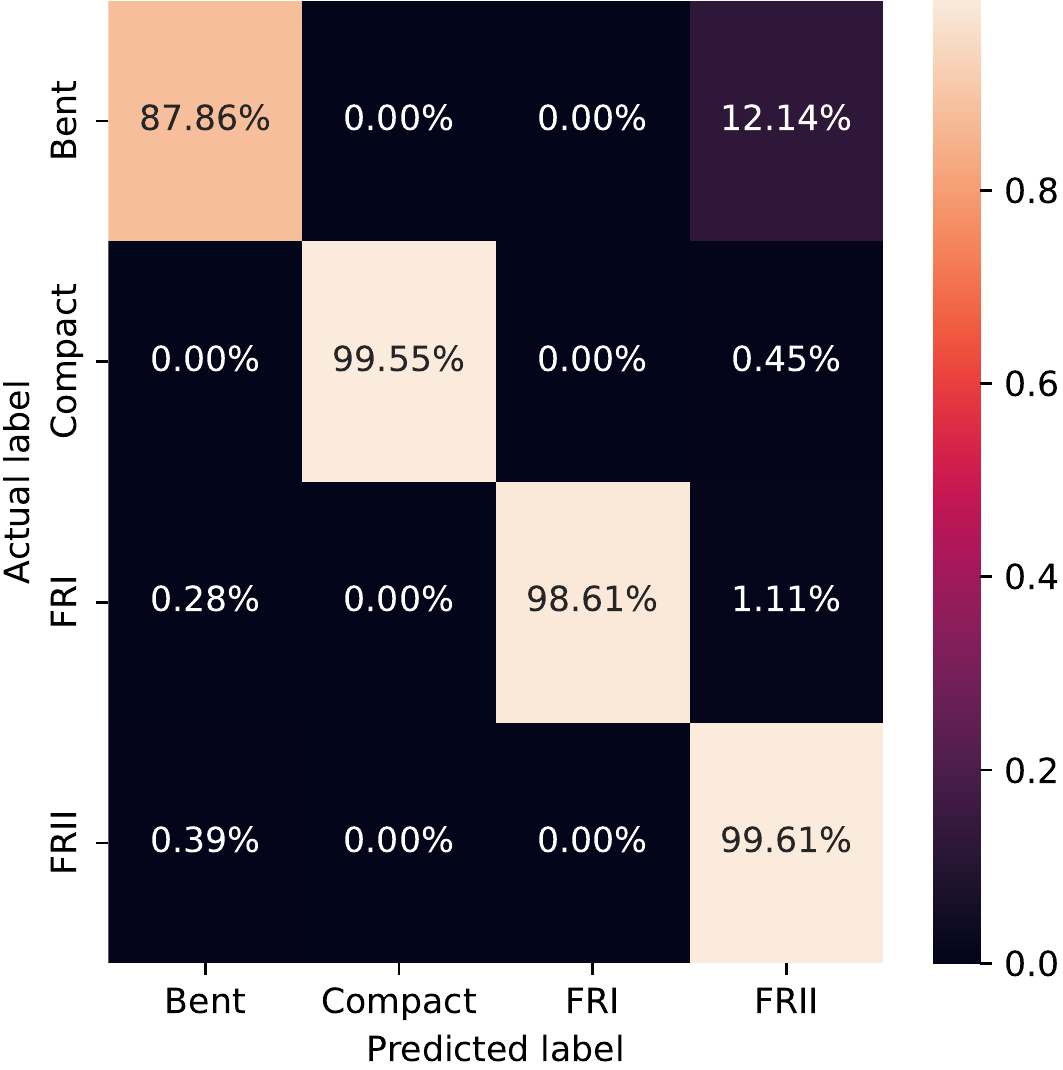}
    \caption{Augmentation}\label{fig:cm_aug}
    \end{subfigure}
    \caption{Average confusion matrices of SCNNs with various rotational preprocessing strategies.}.
    \label{fig:confusion}
\end{figure*}

Upon further investigation of the confusion matrices shown in \autoref{fig:confusion} and the per-class metrics shown in \autoref{tab:rotate_class_results} , we were also able to establish that these improvements in classification results occurred in all four classes. However, it became clear that addressing rotational variations had a considerably larger impact on the classification results of bent and FRII galaxies. It was observed that CNNs were able to achieve good classification performance on compact and FRI galaxies, even when rotational variations were not addressed. This indicates that CNNs are less sensitive to rotations in these galaxies than rotations in bent and FRII galaxies. We hypothesize that this is due to the simpler, compact nature of the FRI and compact galaxies.

\begin{table*}
	\centering
	\caption{Average per-class performance of SCNNs with various rotational preprocessing strategies.}
    \label{tab:rotate_class_results}
	\begin{tabular}{llllllllllllllll}
		\hline
		& \multicolumn{3}{c}{Bent} && \multicolumn{3}{c}{Compact} && \multicolumn{3}{c}{FRI} && \multicolumn{3}{c}{FRII}\\
        \cline{2-4}\cline{6-8}\cline{10-12}\cline{14-16}
        Preprocessing & Precision & Recall & F1 & & Precision & Recall & F1 && Precision & Recall & F1 && Precision & Recall & F1\\
		\hline
        None & 0.8 & 0.695 & 0.743 && 0.937 & 1.0 & 0.966 && 0.983 & 0.894 & 0.936 && 0.823 & 0.899 & 0.858\\
        Standardisation & 0.961 & 0.898 & 0.928 && 0.963 & 1.0 & 0.981 && 0.975 & 0.922 & 0.947 && 0.924 & 0.968 & 0.945\\
        Augmentation & 0.99 & 0.879 & 0.93 && 1.0 & 0.995 & 0.998 && 1.0 & 0.986 & 0.993 && 0.932 & 0.996 & 0.963\\
		\hline
	\end{tabular}
\end{table*}

Furthermore, it can be determined from the results in \autoref{tab:rotate_results} and \autoref{tab:rotate_class_results} that rotational augmentation also leads to better classification performance than rotational standardisation. That being said, \autoref{tab:rotate_class_results} indicates that rotational standardisation results in a slightly better recall for bent and compact galaxies, which indicates that CNNs that were trained and evaluated on standardised galaxies are marginally better at identifying these classes.

Upon closer inspection of the samples on which classification errors are made, we found that there was an overlap between the models using augmentation and standardisation. However, there were also a number of samples on which only standardisation or augmentation led to errors. This could indicate that standardisation and augmentation are relying on different features or have different vulnerabilities. Unfortunately, we could not identify any clear similarities between samples on which each technique tended to make mistakes, nor could we establish any clear differences between samples on which only standardisation models would make a mistake and samples on which only augmentation models would make a mistake.

\begin{figure*}
    \centering
	\begin{subfigure}{.35\linewidth}
    \includegraphics[width=\linewidth]{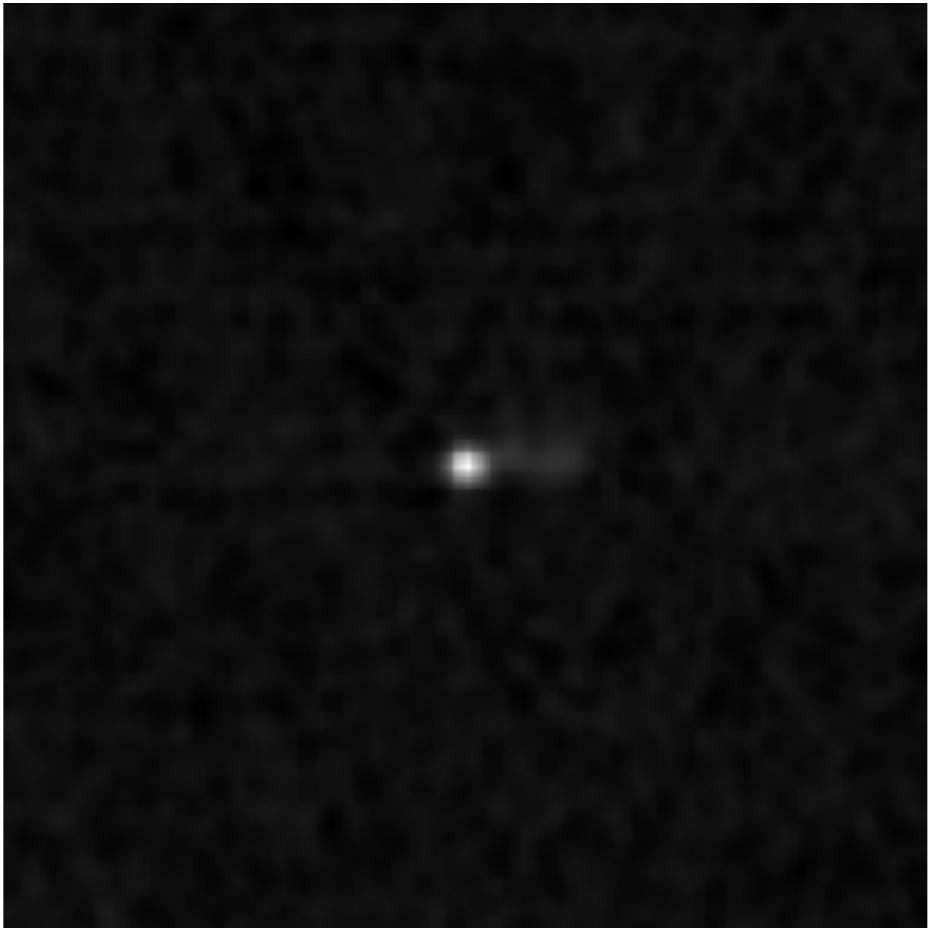}
    \caption{Example of a FRI sample where thresholding produced a background artefact.}\label{fig:rot_fail_thresh_before}
    \end{subfigure}\hspace{5mm}
    \begin{subfigure}{0.35\linewidth}
    \includegraphics[width=\linewidth]{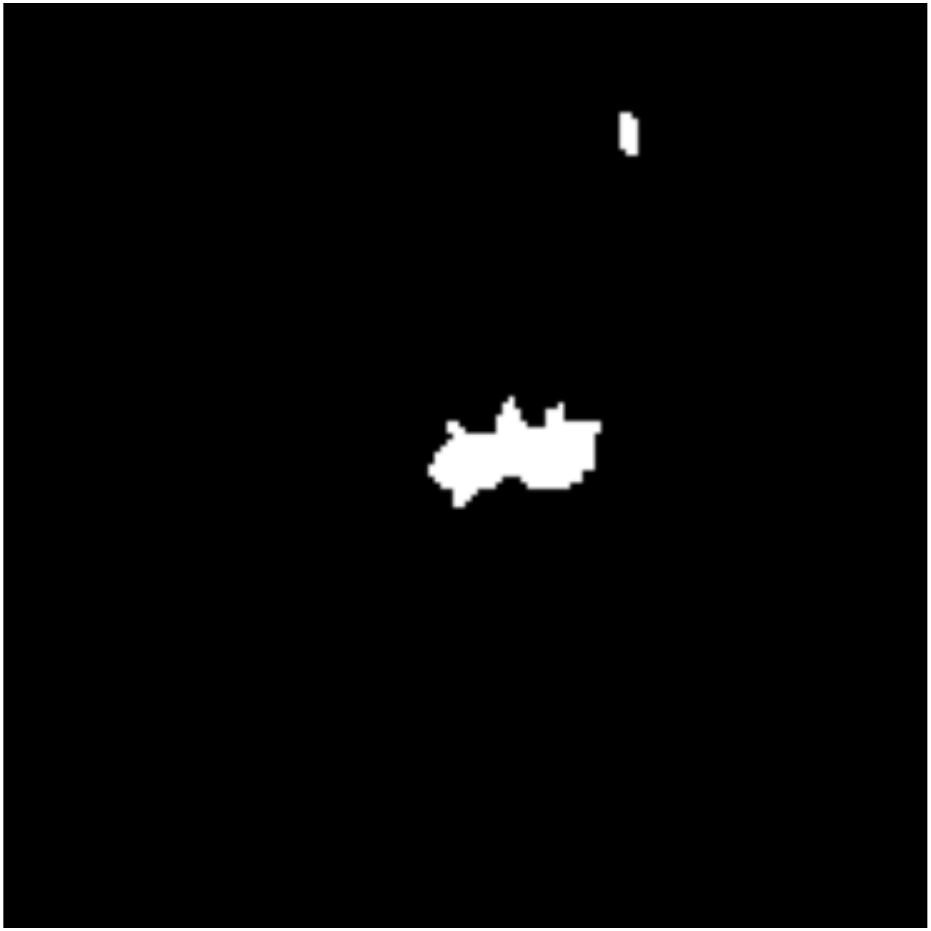}
    \caption{The thresholding mask that corresponds to \autoref{fig:rot_fail_thresh_before}. Notice the artefact in the top right corner.}\label{fig:rot_fail_thresh_after}
    \end{subfigure}
    \begin{subfigure}{0.35\linewidth}
    \includegraphics[width=\linewidth]{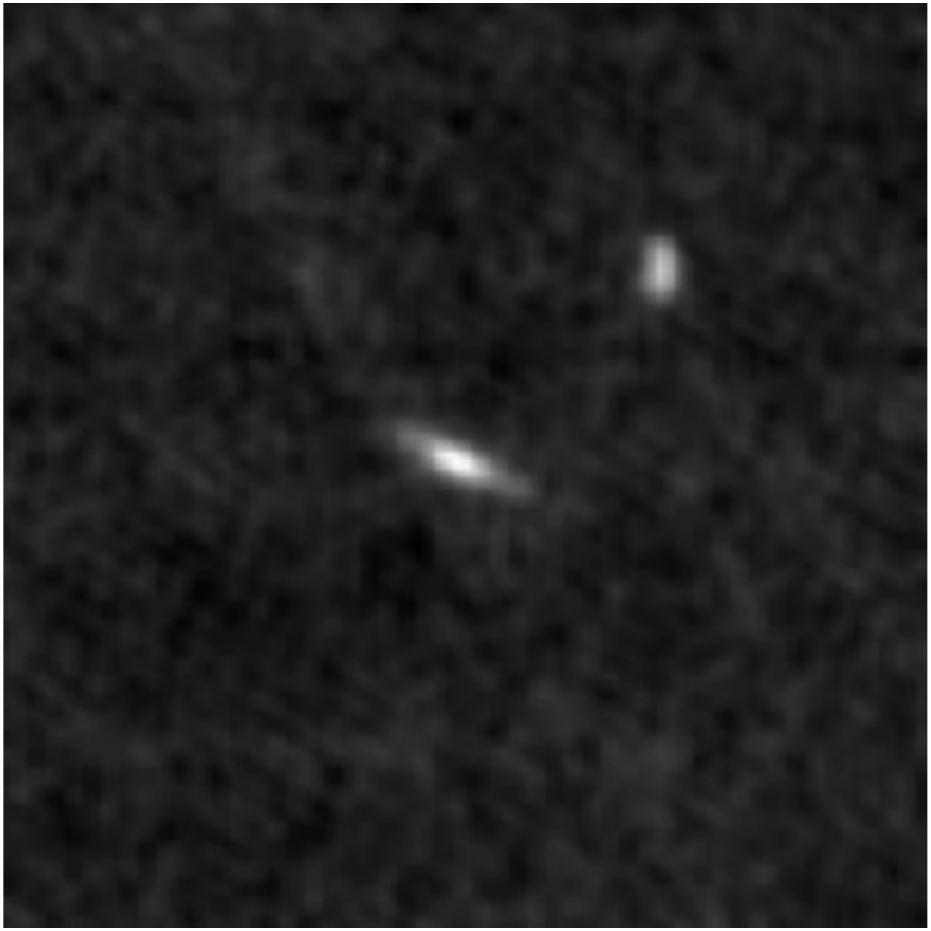}
    \caption{Example of an additional radio source in the background of a FRI sample.}\label{fig:rot_fail_source}
    \end{subfigure}
    \caption{Samples where rotational standardisation failed due to background elements.}.
    \label{fig:rot_fail}
\end{figure*}

We suspect that the reason why augmentation generally leads to a better F1 score than rotational standardisation is due to the sensitivity of rotational standardisation to the presence of additional radio sources and thresholding artefacts in the background. To clarify what this means, we have included examples in \autoref{fig:rot_fail}. These sources and artefacts lead to changes in the principal components, which in turn leads to incorrect derotations.

To further investigate the impact of these incorrect derotations, we took a look at their prevalence in our dataset. What we found is that 32 samples were incorrectly derotated due to additional radio sources and 6 samples were incorrectly rotated due to thresholding artefacts. Respectively, this corresponds to approximately $3.6\%$ and $0.6\%$ of the total dataset. When looking at the class distribution of these incorrect derotations, as shown in \autoref{tab:failed_rots}, we can see that these problematic samples only affect the bent, FRI and FRII classes and that they form a particularly large percentage of the FRI and bent samples. The incorrectly derotated samples might not always lead directly to classification errors, but they will have an indirect effect on the models' ability to classify samples from that class. The models will struggle to extract useful information from these problematic samples, because the information is at a different orientation, which will lead to a loss of training information in an already small dataset. It is also entirely possible that these samples will lead to confusion within the models, seeing as they will differ considerably from the other samples that have been standardised and thus increase the intra-class variance. Our suspicions seem to be confirmed by the correlation between the large portion of problematic FRI samples and the large difference in FRI classification performance between data augmentation and rotational standardisation that we can observe in \autoref{tab:rotate_class_results} and \autoref{fig:confusion}. The difference in classification performance for the FRII galaxies is also considerably smaller, which corresponds with the smaller percentage of samples that were incorrectly derotated. However, we should note that this correlation between the difference in classification performance and the percentage of incorrectly derotated samples does not hold for bent galaxies. We suspect that this is due to the fact that bent galaxies have no clear correct orientation, which means that the "incorrect" derotations will only lead to a marginal increase in intra-class variance.

\begin{table*}
	\centering
	\caption{Distribution of incorrect derotations.}
    \label{tab:failed_rots}
	\begin{tabular}{lll}
		\hline
        Class & Total Incorrect Derotations & Percentage of Class\\
		\hline
		Bent & $18$ & $8.5\%$\\
        Compact & $0$ & $0.0\%$\\
		FRI & $17$ & $9.3\%$\\
        FRII & $12$ & $3.2\%$\\
		\hline
	\end{tabular}
\end{table*}

We hypothesize that if one were to fine-tune the thresholding algorithm to remove these background radio sources, rotational standardisation would be able to compete with rotational augmentation in terms of classification performance.

That being said, the true advantage of rotational standardisation lies in the amount of time required to train the networks. The average training time for CNNs that were trained on augmented data is almost a factor of 5 times longer than the times required to train CNNs on samples with a standardised orientation. The training times for augmented data will become even longer if the rotational intervals are reduced. This confirms our hypothesis that data augmentation will be considerably slower due to the consideration of dataset size.

What is interesting to note, is that on average, the CNNs also trained faster on the samples with a standardised orientation than the samples where no rotational preprocessing was applied. This result could be due to the fact that conventional CNNs have to learn what a feature looks like at various orientations, as was explained by \citet{Scaife2021}. Thus, if rotational variations are reduced or removed, it is highly likely that less time will be spent on learning these duplicate feature filters, which in turn means that the CNN will be able to learn a variety of informative features in less time.

This reduction in training time is likely to become more prominent as the complexity of the neural networks increases or if larger datasets are used that include more samples or more classes. Thus, in time-sensitive scenarios it might be beneficial to rather use rotational standardisation instead of rotational augmentation.

\subsection{Performance of Guided Neural Networks on XOR}
\label{sec:guided_xor_results}
The reader might recall that a variant of the merged architecture was evaluated on the XOR dataset first to determine whether it was feasible to guide the training of neural networks. For more information regarding the architecture, the reader is referred to Appendix~\ref{appendix:XOR_arch}.

\begin{figure}
    \centering
    \includegraphics[width=\linewidth]{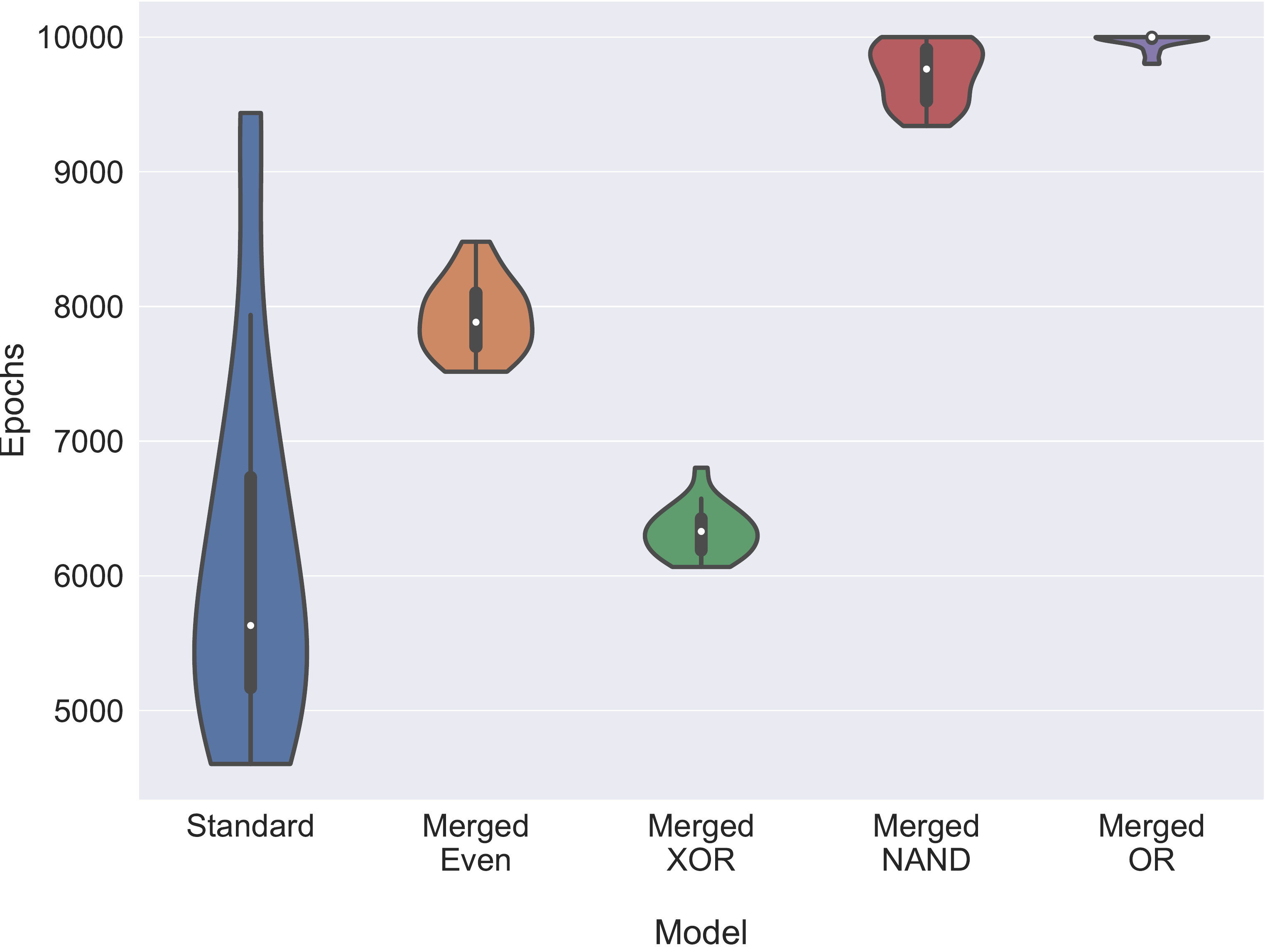}
    \caption{Epochs taken to train the XOR networks. The given labels indicate the network architecture, as well as the weight distribution that was used for each merged network. The reader is referred to Appendix~ \ref{appendix:XOR_arch} for clarity on the meaning of the labels.}
    \label{fig:xor_box}
\end{figure}

In \autoref{fig:xor_box}, one can see violin plots representing the number of training epochs it took for the XOR neural networks to reach a loss smaller than 0.01. These violin plots have a box plot at their center. The thin lines in the box plot indicate the values smaller than the 25th quantile and larger than the 75th quantile, the thick line indicates the values that are between the 25th and 75th quantile and the white dot indicates the median value. Surrounding this box plot is a density estimation that is based on the observed values. This assists in visualising the expected distribution of each of the performance metrics. The wider the density estimation is, the more likely we are to observe the values in that region. For this paper, the density estimations were restricted to the range of values observed in our experiments, thus no estimation is done for values outside of the range of observed values. Estimations occurring outside of the box plot is due to the fact that the box plots exclude values that are deemed to be outliers, but the density estimation does not.

These violin plots indicate a large difference between the training process of the guided and unguided XOR neural networks. The training process of the guided networks is clearly much stabler. We can see that all four of the guided networks have a very compact violin plot, whilst the unguided network has a violin plot that spans quite a large range of epochs. This is to be expected, seeing as the guided networks will look for similar patterns during each training run, whilst the unguided network is more likely to explore various different patterns in each run.

We can also observe that the merged network with a heavy XOR weight is clearly the guided network that trains the quickest. This makes sense, seeing as the other guided networks might spend too much time on fine tuning the results of the NAND and OR neurons. Another observation that can be made is that the standard XOR network will train quicker than the guided networks in most runs. However, it can also train considerably longer than some of the guided networks, such as the merged network with a heavy XOR weight.

\begin{figure}
    \centering
    \includegraphics[width=\linewidth]{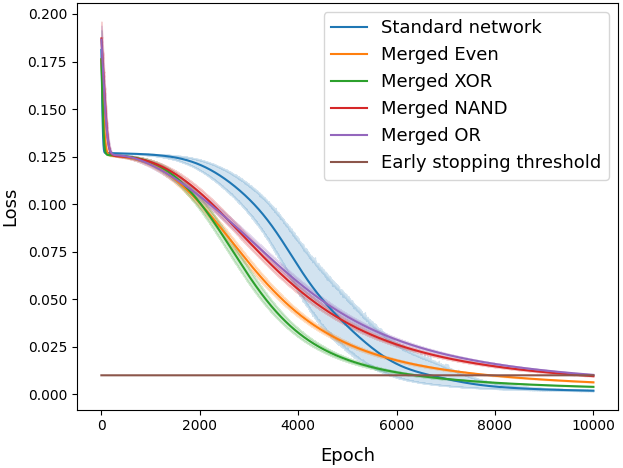}
    \caption{Loss curves of XOR networks during training.}
    \label{fig:xor_loss}
\end{figure}

To investigate why these results were observed, the loss curves of each of the neural networks were plotted after training. These curves can be seen in \autoref{fig:xor_loss}. The reader should note that these loss curves do not correspond to the training runs for the violin plots, seeing as early stopping had to be disabled to record the entire loss curve over 10000 epochs. In these loss curves the solid line represents the average loss at each epoch and the shaded region indicates the 95\% confidence interval. The horizontal line at the bottom of the figure indicates the early stopping threshold that was used.

These loss curves also indicate that the training process of the merged neural networks is much stabler than that of the standard neural network, seeing as their confidence intervals are considerably more compact. Furthermore, these loss curves indicate that the merged networks perform better than the violin plots suggested. These loss curves suggest that all of the merged neural networks would outperform the standard neural network if the threshold of early stopping was increased. It should also be noted that the merged neural networks will outperform the standard neural network if training was conducted for less epochs. For example, if training was only conducted for 3000 epochs, all of the merged networks would achieve a better loss than the standard network.

Merged neural networks seem to converge much quicker initially, but the convergence slows down in later epochs. This behaviour makes sense, seeing as the merged networks will immediately start learning to extract features and patterns that have been deemed to be informative, whilst the standard neural network will spend more time exploring the search space and is thus likely to take longer to find good features. The quick convergence of these merged networks could be very beneficial for time-sensitive applications.

\subsection{Performance of Guided CNNs on FRGMRC Data}
\label{sec:guided_results}

\subsubsection{Algorithmically extracted features}
\label{sec:guided_galaxy_results}
\begin{figure}
    \centering
    \includegraphics[width=\linewidth]{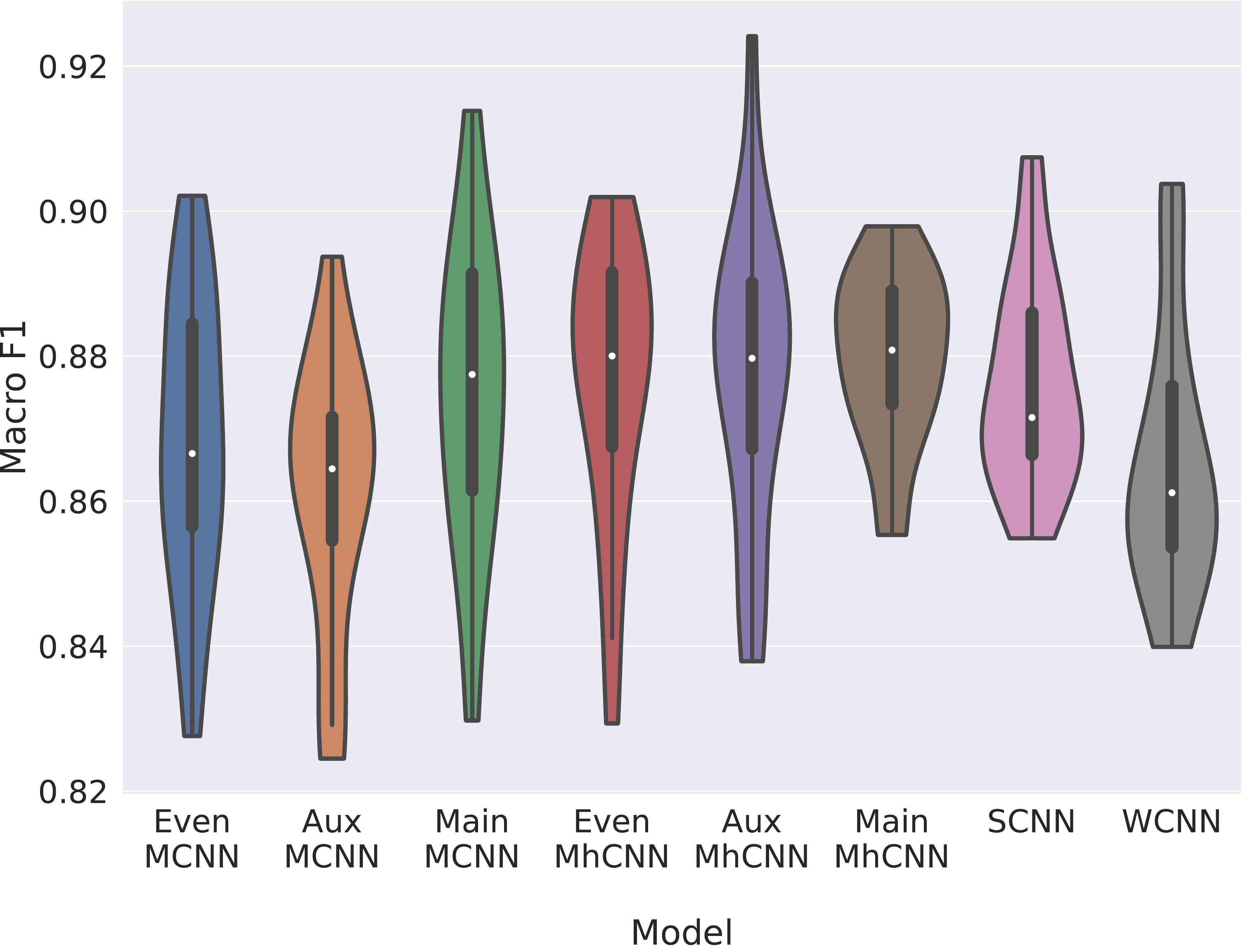}
    \caption{Violin plots of the macro F1 scores from Section~\ref{sec:f1_score} that were observed when evaluating models on the test dataset. The reader is referred back to \autoref{tab:weights} for an overview of the weights used in the MhCNNs and MCNNs.}
    \label{fig:auto_galaxy_f1}
\end{figure}

\begin{figure}
    \centering
    \includegraphics[width=\linewidth]{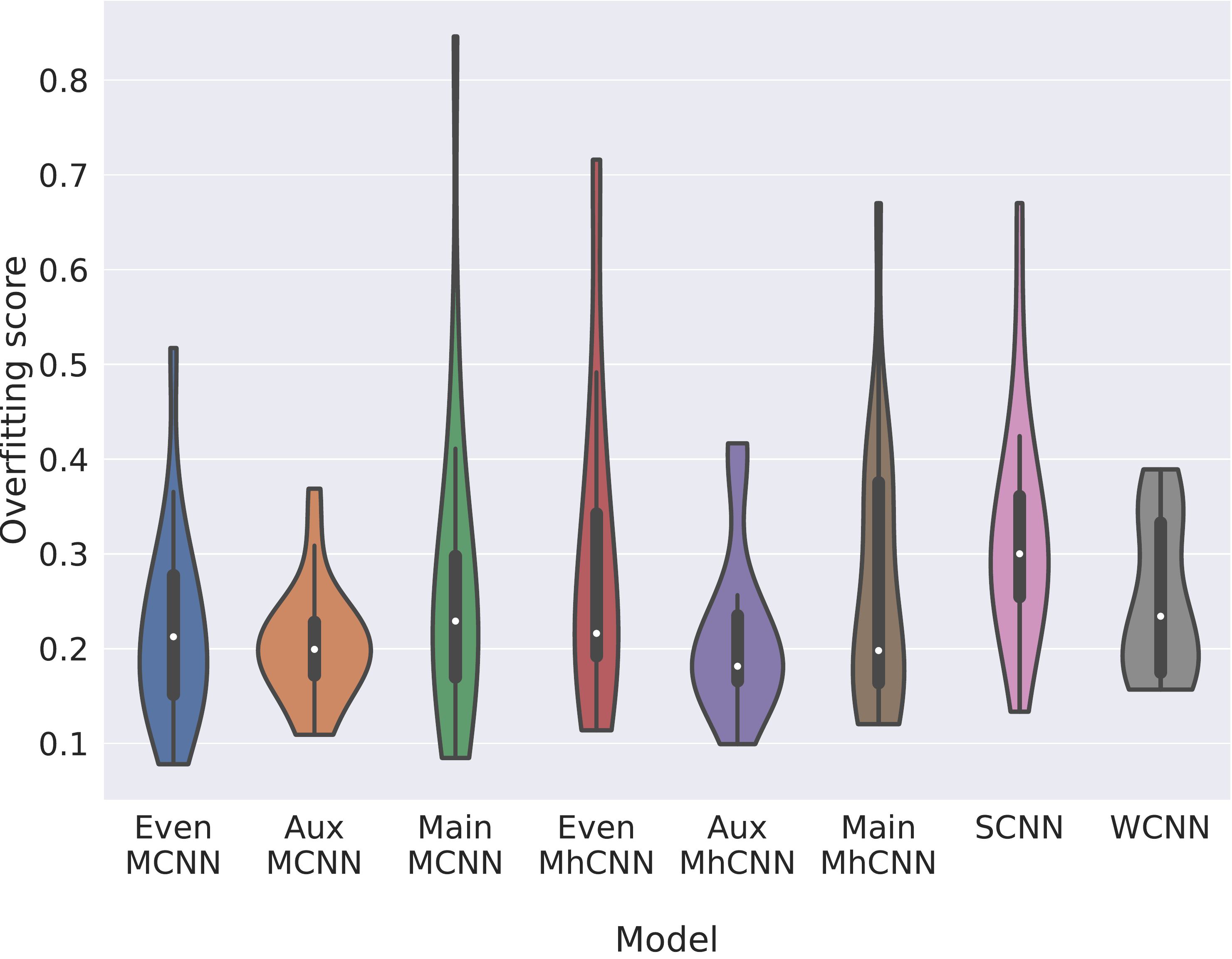}
    \caption{Violin plots of the overfitting scores from Section \ref{sec:overfit_metric} that were observed after training the models.}
    \label{fig:auto_galaxy_overfit}
\end{figure}

\begin{figure}
    \centering
    \includegraphics[width=\linewidth]{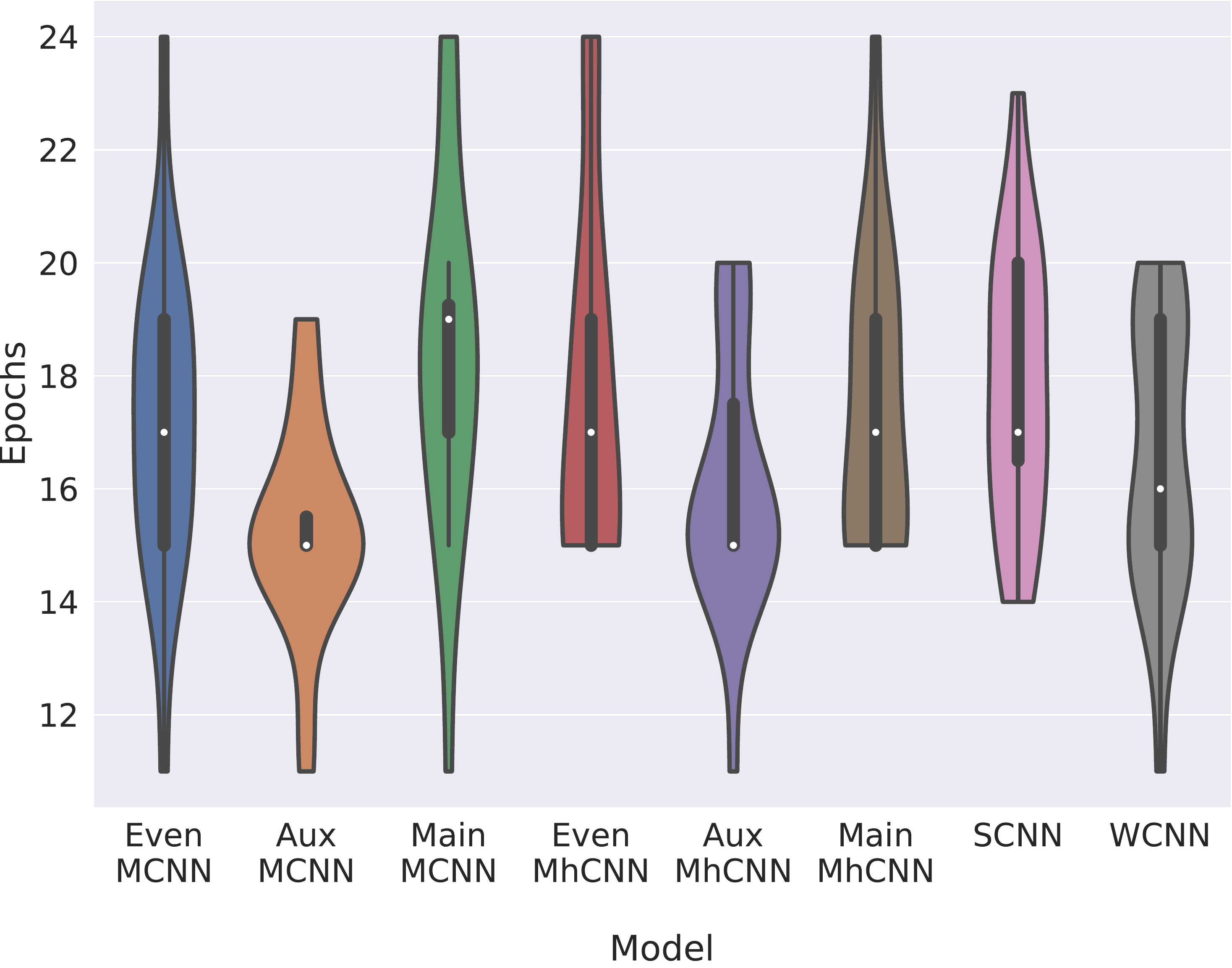}
    \caption{Violin plots of the number of epochs it took to train the CNNs.}
    \label{fig:auto_galaxy_epochs}
\end{figure}

To provide an accurate depiction of the performance of the CNNs on the FRGMRC data, we generated a few violin plots that present the observed values of the performance metrics for each run of each of the CNNs. We will focus on the violin plots of the macro F1 scores, the overfitting metric and the number of epochs needed to train the networks, seeing as they were the most informative when evaluating the performance of the CNNs. The plots are shown in \autoref{fig:auto_galaxy_f1}, \autoref{fig:auto_galaxy_overfit} and \autoref{fig:auto_galaxy_epochs} respectively.

When comparing the white dots that represent the median F1 score in \autoref{fig:auto_galaxy_f1} we can observe that all of the MhCNNs, as well as the MCNN with a heavier weight for the main cross-entropy loss managed to achieve a better macro F1 score when classifying each of the radio galaxies. However, these scores are only marginally better than that of the SCNN.

Shifting our focus to \autoref{fig:auto_galaxy_overfit}, the benefits of the guided networks become more apparent. We can immediately observe that overfitting is less likely to occur in the MCNNs and MhCNNs that use a larger weight for the auxiliary loss component. We can also observe a similar trend in the WCNN.

These networks all have something in common, they make use of higher quality features to guide the classification of the samples. To understand why this is true for the merged and multiheaded networks, the reader is reminded that these networks made use of a considerably larger weight for the loss that corresponds to the features that are extracted from the samples. During training, the feature extraction will thus be optimized much quicker than the classifications, which means that the network will have access to high quality features much sooner than the other guided networks. The wide network will also have high quality features, because they were extracted by a pretrained ACNN.

It is very likely that it is these high quality features that makes the networks more robust to overfitting. We suspect that the networks are more likely to learn to use these features for classification if they are trustworthy at an early stage of training. If the network makes use of these features, they will not need to find as many other features in the training data, which reduces the risk of overfitting on noise or uninformative features.

From \autoref{fig:auto_galaxy_epochs} it also becomes apparent that the auxiliary-weighted MCNNs and MhCNNs, as well as the WCNNs generally require less epochs to train. This is likely due to similar reasons as the reduction in overfitting. Due to the presence of high quality feature vectors early in the training process, we suspect that these networks don't need to spend as much time searching for features that assist in classifying the given samples.

The reader should note that, as shown by \autoref{tab:galaxy_times}, some of the networks don't necessarily train quicker if they train in less epochs\footnote{The training time of the WCNNs does not include the time taken to train the ACNNs. This was done to show what the training time would be if these features were already available.}. The MCNNs train in less epochs and less time than the SCNNs, but the MhCNNs have a much more complex structure with more parameters to tune than the other CNNs, which means each epoch takes longer to complete. However, the fact that they train in less epochs still indicates that by guiding the networks to look for specific features, we can help the networks to converge in less training steps. Furthermore, if the code is optimised to train the two heads of the MhCNNs in parallel, it is likely that one will also be able to train the MhCNNs in less time than the SCNNs.

\begin{table}
	\centering
	\caption{Average training times of CNNs on galaxy dataset.}
    \label{tab:galaxy_times}
	\begin{tabular}{ll}
		\hline
		Network & Average Training Time (s)\\
		\hline
		SCNN & $116$\\
		WCNN & $99$\\
		Even MhCNN & $144$\\
		Aux MhCNN & $127$\\
		Main MhCNN & $137$\\
		Even MCNN & $119$\\
		Aux MCNN & $104$\\
		Main MCNN & $119$\\
		\hline
	\end{tabular}
\end{table}

What is also interesting to note is that the box plots of the auxiliary-weighted MCNNs and MhCNNs is much more compact than that of the standard network, which is similar to the behaviour that we observed for the XOR dataset. This is especially true for the box plot of the MCNNs. The compact box plots could be another indication that the networks spend less time searching for features and more time optimizing the extraction and use of the features that we proposed. This leads to less variability in the training process and thus the number of epochs required for training becomes more consistent.

We suspect that the MCNNs have a more compact box plot than the MhCNNs, because the selected features play a more direct role in the classification of the samples. Thus, if the selected features are truly informative and the network is able to accurately extract them, there is no need for the network to look for other features. Due to the indirect role that the selected features play in the MhCNNs, it might be necessary for the MhCNNs to spend more time to explore the search space. Seeing as the convolutional layers have been trained to also extract the selected features, the chances are good that the MhCNN will still find these features and make use of them for classification, but it will take more time. The networks are also likely to find other features in the process, which will lead to an increase in the range of epochs that are required for training.

\subsubsection{Manually extracted features}
\begin{figure}
    \centering
    \includegraphics[width=\linewidth]{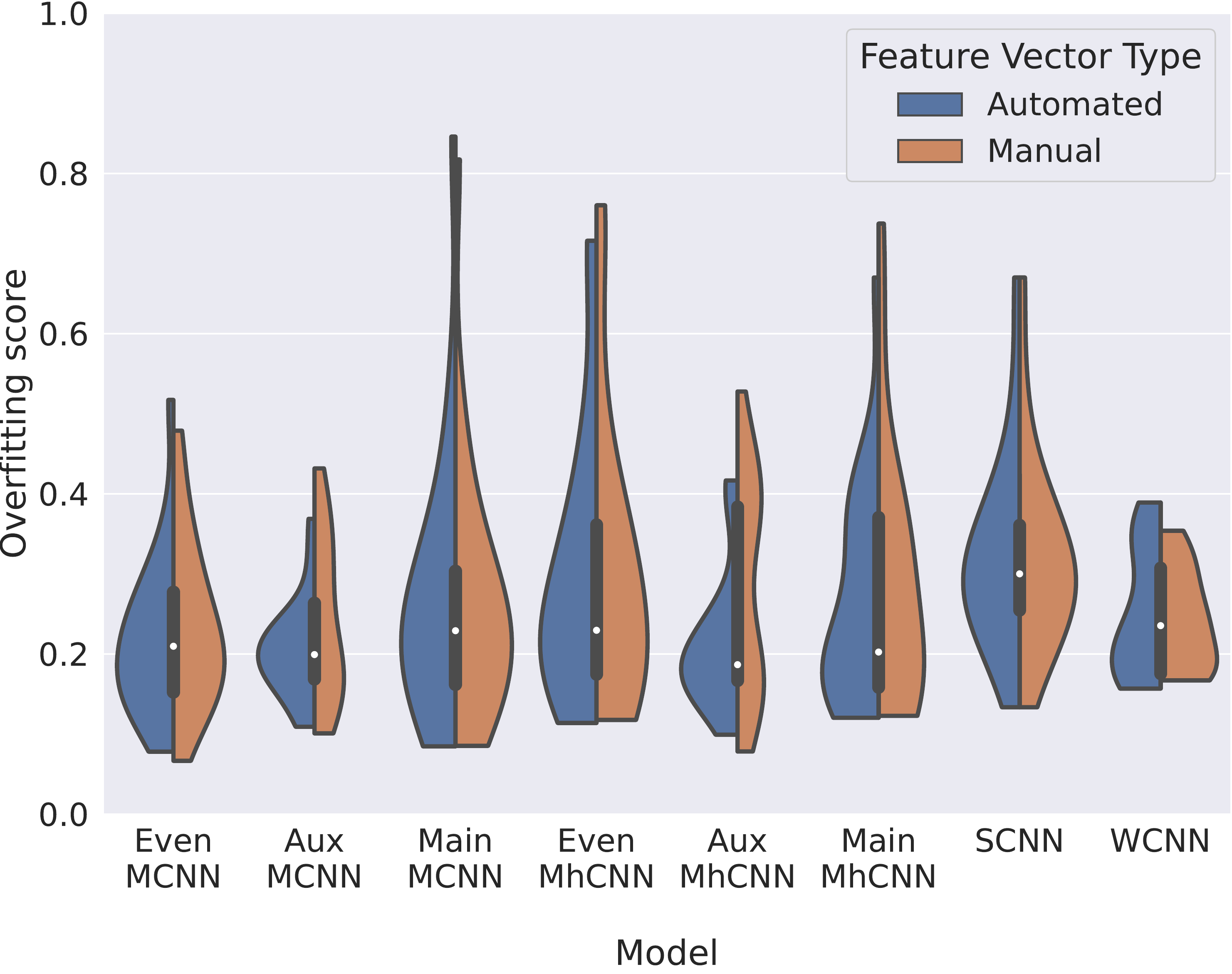}
    \caption{Comparison of overfitting when using manual and algorithmically extracted features.}
    \label{fig:comp_galaxy_overfit}
\end{figure}

\begin{figure}
    \centering
    \includegraphics[width=\linewidth]{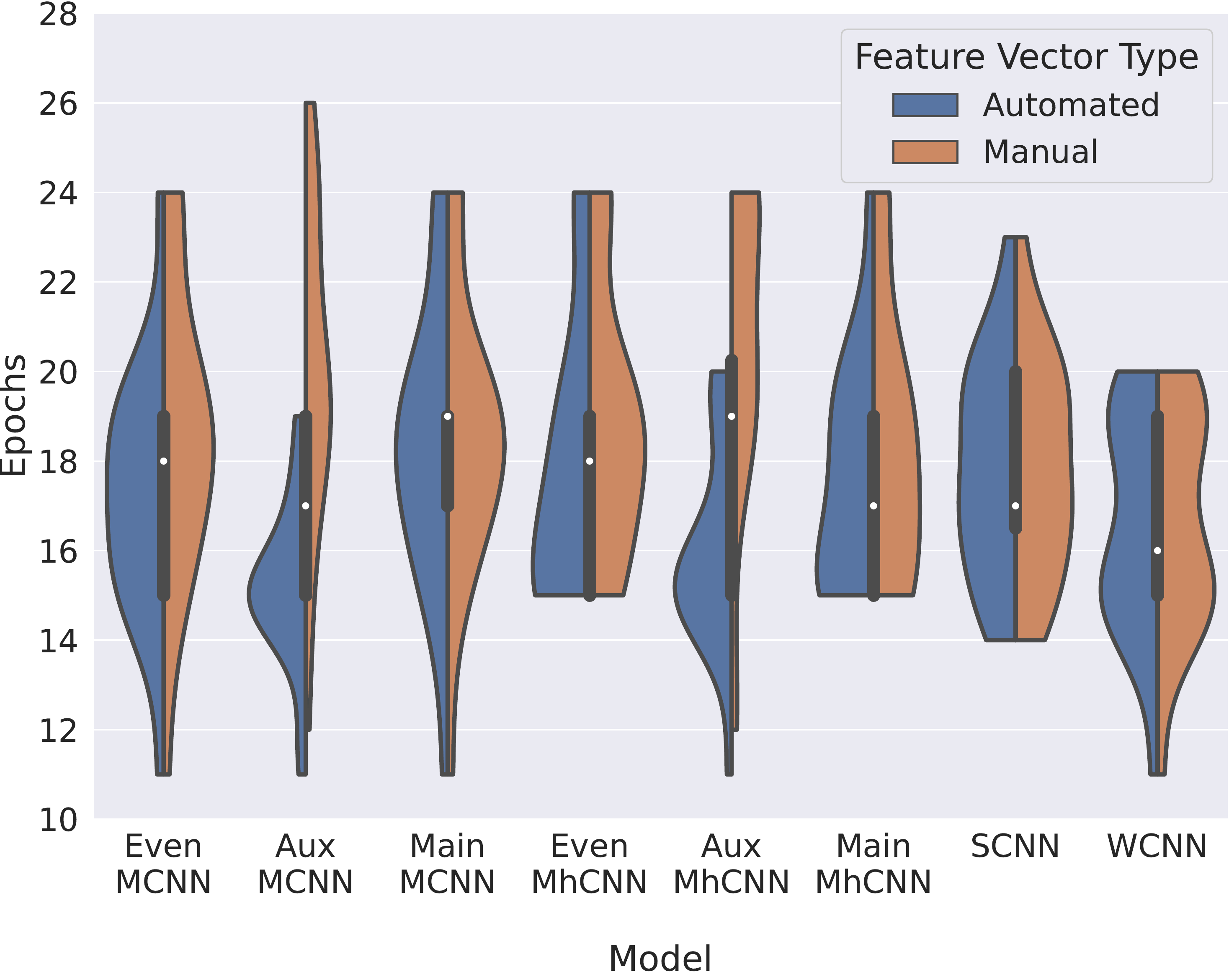}
    \caption{Comparison of epochs taken when using manual and algorithmically extracted features.}
    \label{fig:comp_galaxy_epochs}
\end{figure}

As has been mentioned, we also have a set of feature labels where some of the features were extracted manually from each sample. We evaluated the guided CNNs on this set of features as well, because we wanted to determine whether it would make a significant difference in the networks' performance. In this section we will mainly focus on the violin plots shown in \autoref{fig:comp_galaxy_overfit} and \autoref{fig:comp_galaxy_epochs}.

From these figures, it becomes apparent that the auxiliary-weighted MhCNNs and MCNNs, as well as the WCNNs, still tend to overfit less. However, the guided networks also seem to perform worse when we make use of the manually extracted features.

This behaviour seems counter-intuitive, but it makes sense after considering that the manually extracted features are not necessarily accurate, especially the core count. The feature that indicates whether the body of the galaxy is curved is likely to be accurate, because it is fairly straightforward to determine if this is true or not when looking at the images. However, there were no clear rules for what should be considered to be a core, which means that the core counts were based on the intuition of the person who labelled them. This could lead to inaccurate core counts, which in turn is likely to lead confusion within the neural networks when they attempt to learn how to extract this feature.

To investigate whether our suspicions were true, we calculated the average MSE of the core counts predicted by the auxiliary-weighted MCNNs after they were trained with the manually and automatically extracted core counts. We found that the MSE was 0.029 when using the automatically extracted core counts and that the MSE was 0.113 when using the manually extracted core counts. This confirms our suspicion that the manually extracted core counts lead to confusion in our guided networks which in turn would negatively affect the overall performance of these networks.

Thus, we suggest that algorithmically extracted features are used, unless one is certain that the features can be accurately extracted manually, using only information that is available in the training data.

\section{Conclusion}
\label{sec:conclusion}
The two goals of this paper were to investigate the benefits of rotational standardisation as well as the benefits of guiding neural networks to look for specific features.

In Section~\ref{sec:rotate_results} it was shown that rotational standardisation definitely leads to better classification performance than if no attempt is made to address rotational variations. Rotational standardisation also managed to achieve this improvement in much less time than rotational augmentation can. That being said, augmentation might still be the better choice in situations where time and computational power is not a problem, seeing as it can achieve better results.

It was also identified that rotational standardisation is susceptible to additional radio sources in the images, as well as thresholding artefacts. Thus, if steps can be taken to address these problems in the images, we are of the opinion that rotational standardisation will be able to rival rotational augmentation, both in training time and classification accuracy.

When evaluating the performance of guided neural networks on the XOR dataset in Section~\ref{sec:guided_xor_results}, we determined that the training process could become stabler and converge much quicker if we guided the neural networks to look for informative features. These results are promising and clearly indicate that the training process can be affected positively by guiding the networks.

After evaluating the guided architectures on the FRGMRC data in Section~\ref{sec:guided_galaxy_results}, we confirmed that it was vitally important that accurate and informative features were used to guide the training process. Furthermore, we confirmed that if these features were accurate and informative, it was possible to reduce overfitting in the networks. It was also possible to reduce the average time needed for the training of the WCNNs as well as the MCNNs with a larger weight for the auxiliary loss. We are of the opinion that these guided architectures will be even more effective if the feature extraction process is fine-tuned.

These results indicate that such guided architectures could be beneficial in time-sensitive applications, in applications where a stable training process is essential, as well as in applications with multiple goals. Specifically, the multiheaded and merged neural networks could be very useful in applications where one wants to identify specific features that could also be informative when classifying samples. As was seen in this paper, these guided neural networks can be trained to achieve two goals in less time than it might take to achieve one, as long as these goals can contribute to one another.

Future work could investigate ways to make the rotational standardisation approach more robust to background information in the images. It could also consist of comparing rotational standardisation to group-equivariant CNNs or constructing new guided architectures.

\section*{Acknowledgements}
MV acknowledges financial support from the Inter-University Institute for Data Intensive Astronomy (IDIA), a partnership of the University of Cape Town, the University of Pretoria, the University of the Western Cape and the South African Radio Astronomy Observatory, and from the South African Department of Science and Innovation's National Research Foundation under the ISARP RADIOSKY2020 Joint Research Scheme (DSI-NRF Grant Number 113121) and the CSUR HIPPO Project (DSI-NRF Grant Number 121291).

MP acknowledges financial support from the Inter-University Institute for Data Intensive Astronomy (IDIA) and from the Center of Radio Cosmology (CRC, NRF Grant Number 84156) at the University of the Western Cape.

\section*{Data Availability}
The FRGMRC and the supporting FIRST fits cutouts used for our work are publicly available at \url{https://doi.org/10.5281/zenodo.7645530}.



\bibliographystyle{mnras}
\bibliography{ms} 



\appendix

\section{XOR Neural Network Architectures}
\label{appendix:XOR_arch}
In this section we present the architectures used for the standard unguided neural networks and the merged guided neural networks that were applied to the XOR dataset. Section~\ref{appendix:XOR_std} will present the standard network and Section~\ref{appendix:XOR_merged} will present the merged network.

\subsection{Standard XOR Neural Network}
\label{appendix:XOR_std}
It would not make sense to apply CNNs to the XOR dataset. These networks are unnecessarily complex for simple tabular data.

\begin{figure}
    \centering
    \includegraphics[width=.9\linewidth]{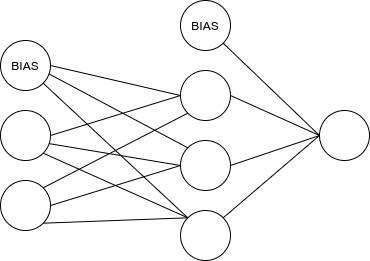}
    \caption{Standard XOR neural network.}
    \label{fig:stdXOR}
\end{figure}

Thus a simple neural network was created for the XOR operator. This network only has an input layer with two input neurons, a hidden layer with three hidden neurons and an output layer with one output neuron. Each neuron makes use of the sigmoid activation function and the MSE variant that was discussed in Section~\ref{sec:mse} was used for the loss function. A representation of the architecture can be seen in \autoref{fig:stdXOR}.

This standard neural network was used as a baseline with which to compare the performance of the guided neural network.

\subsection{Merged XOR Neural Network}
\label{appendix:XOR_merged}
\begin{figure}
    \centering
    \includegraphics[width=1\linewidth]{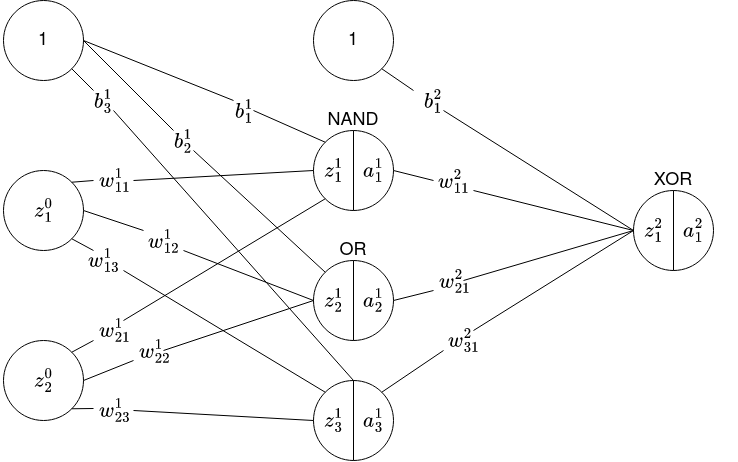}
    \caption{Merged XOR neural network.}
    \label{fig:mergedXOR}
\end{figure}

A representation of the merged network for the XOR operator can be seen in \autoref{fig:mergedXOR}. In this figure $b$ represents bias terms, $w$ represents the weight terms, $z$ represents neuron inputs before applying the activation function and $a$ represents the inputs after applying the activation function.

As can be seen in this figure, two of the hidden neurons are guided to represent the NAND and OR binary operators, which are then fed back into the output neuron that represent the XOR operator. To understand why this was done, consider the fact that the XOR operator can be rewritten in terms of the NAND and OR operators: $x_1$ XOR $x_2 = (x_1$ NAND $x_2)$ AND $(x_1$ OR $x_2)$.

To ensure that gradient descent does not only optimize the XOR function, a loss function was used that aggregates the error of the XOR, NAND and OR outputs. This loss function simply calculates the weighted sum of each output's mean squared error. This weighted sum can be seen in \autoref{eq:xor_loss}.

\begin{equation}
    C = \frac{l_1}{2}(y_{1} - a^2_1)^2 + \frac{l_2}{2}(y_{2} - a^1_1)^2 + \frac{l_3}{2}(y_{3} - a^1_2)^2
	\label{eq:xor_loss}
\end{equation}

The weights that were used for this loss function can be seen in \autoref{tab:xor_weights}.

\begin{table}
	\centering
	\caption{Weights used to balance loss components in XOR networks.}
    \label{tab:xor_weights}
	\begin{tabular}{llll}
		\hline
		Network & XOR weight ($l_1$) & NAND weight ($l_2$) & OR weight ($l_3$)\\
		\hline
		Even merged & 0.34 & 0.33 & 0.33\\
        XOR merged & 0.5 & 0.25 & 0.25\\
        NAND merged & 0.25 & 0.5 & 0.25\\
        OR merged & 0.25 & 0.25 & 0.5\\
		\hline
	\end{tabular}
\end{table}

\subsection{Generalisability of guided architectures.}
When guided architectures encounter sources with different feature variations or new features, there might be a concern that they will fail to generalise and will perform poorly. In this section we will shortly discuss this concern and our suggested approach to address it.

The first scenario that one should consider is the presence of new features in our samples, whether in training or in inference. If we encounter new informative features during training, our guided architectures should be able to learn what these features are and how to extract them by using the free neurons within the network, just as a standard CNN would. However, if we only encounter these new features during inference, it is unlikely that our guided architectures will be able to sufficiently utilize them, but this would also hold true for standard CNNs. CNNs in general cannot utilize features during inference that they have not observed during training.

Another scenario to consider would be rare or unusual variations of the galaxies' features. We are of the opinion that the guided architectures will not be affected any worse by these variations than standard architectures, seeing as all of these architectures are unlikely to be able to learn much from these variations if they are rare or uncommon in our datasets. If, however, we have a new dataset where these feature variations are no longer rare or unusual, our guided architectures will be able to learn from them by making use of the free neurons. That being said, we do acknowledge that the guided architectures do not attribute as much computational power to finding new features as the standard architectures do. Thus, if the change in feature variations is so severe that the features we selected are no longer informative, it is possible that our guided architectures will perform worse than standard CNNs. We should consider, however, whether we expect such severe changes in our feature space and whether these changes would not represent an entirely new class of radio galaxies instead of an unusual instance of a known class.

To address the concern that we might encounter such unexpected or rare feature variations in future large-scale surveys, we suggest the use of an anomaly detection, or more specifically, a novelty detection system that is trained alongside the classification networks. Such a system could be used in parallel with the classification networks during inference to detect when we encounter rare or new feature variations. This system could assist us in finding samples that could prove to be problematic for our guided architectures. If these samples are rare, we can simply extract them and make use of a human expert to determine what they are and how they affected our classification network. However, if we start to encounter such samples on a frequent basis, it could be an indication of concept drift or that we might have found a new class, in which case we would know that it is a good time to re-evaluate and retrain our models.

\bsp	
\label{lastpage}
\end{document}